\documentclass[preprint,preprintnumbers,amsmath,amssymb,nofootinbib,APS]{revtex4}

\usepackage{dcolumn}
\usepackage{bm}
\usepackage{amsmath}
\usepackage{amsfonts}
\usepackage{amssymb}
\usepackage{hyperref}
\usepackage{latexsym}
\usepackage[dvips]{graphicx}
\usepackage{epsf}

\allowdisplaybreaks

\newcommand{\be}{\begin{equation}}
\newcommand{\ee}{\end{equation}}
\newcommand{\bea}{\begin{eqnarray}}
\newcommand{\eea}{\end{eqnarray}}

 \let\b=\beta    \let\d=\delta
     \let\th=\theta   \let\l=\lambda
\let\m=\mu    \let\n=\nu          \let\r=\rho 
\let\s=\sigma      
    
\let\G=\Gamma \let\D=\Delta   \let\L=\Lambda 
         
  \let\eps=\epsilon

\let\==\equiv

\newcommand{\tr}{{\rm tr}}
\newcommand{\Tr}{{\rm Tr}}

\newcommand{\cO}{\mathcal{O}}
\newcommand{\cR}{\mathcal{R}}
\newcommand{\cS}{\mathcal{S}}

\newcommand{\Pt}{\tilde{\Phi}}

\newcommand{\unit}{{\bf{1}}}

\newcommand{\half}{\tfrac{1}{2}}

\newcommand{\cb}{\bar{c}}
\newcommand{\gb}{\bar{g}}

\newcommand{\I}{\mathrm{i}}
\newcommand{\e}{\mathrm{e}}
\newcommand{\p}{\partial}

\newcommand{\Cb}{\bar{C}}
\newcommand{\Db}{\bar{D}}

\newcommand{\Rb}{\bar{R}}

\newcommand{\eb}{\bar{\eta}}

\newcommand{\rmd}{{\rm d}}

\newcommand{\dg}{{\dot{g}}}

\newcommand{\gt}{\tilde{g}}

\newcommand{\tU}{{\tilde{\Upsilon}}}
\newcommand{\tC}{{\widetilde{C}}}

\newcommand{\U}{\Upsilon}

\newcommand{\bQ}{{\bf{Q}}}

\begin{document}
%------------------------------------------------------------------------------

\preprint{PI-QG-120}
\preprint{ITP--UU--09/08}
\preprint{SPIN--09/8}
\preprint{IPhT-T09/024}

\title{\vspace*{4mm}{\bf Taming perturbative divergences in asymptotically safe gravity \\[3ex]}}

\author{Dario Benedetti$^1$}\email{dbenedetti@perimeterinstitute.ca}
\author{Pedro F. Machado$^2$}\email{p.f.machado@uu.nl}
\author{Frank Saueressig$^3$}\email{Frank.Saueressig@cea.fr}
\affiliation{ {\footnotesize {$^1$Perimeter Institute for Theoretical Physics, \nolinebreak
31 Caroline St. N, N2L 2Y5; Waterloo ON, Canada}\\
$^2$Institute for Theoretical Physics; Utrecht University, 3508 TD Utrecht, The Netherlands\\
$^3$Institut de Physique Th\'eorique; CEA Saclay; F-91191 Gif-Sur-Yvette Cedex, France \\
CNRS URA 2306; F-91191 Gif-Sur-Yvette Cedex, France \\
}}

\begin{abstract}
\noindent
We use functional renormalization group methods to study gravity minimally coupled to a free scalar field. This setup provides the prototype of a gravitational theory which is perturbatively non-renormalizable at one-loop level, but may possess a non-trivial renormalization group fixed point controlling its UV behavior. We show that such a fixed point indeed exists within the truncations considered, lending strong support to the conjectured asymptotic safety of the theory. In particular, we demonstrate that the counterterms responsible for its perturbative non-renormalizability have no qualitative effect on this feature.
\end{abstract}

%\pacs{}

\maketitle

%------------------------------------------------------------------------------

%------------------------------------------------------------------------------
\section{Introduction}
%------------------------------------------------------------------------------

Quantized general relativity is notoriously non-renormalizable at the perturbative level.
Such an understanding has been achieved after a number of celebrated calculations that, 
starting with 't~Hooft and Veltman's seminal work \cite{'tHooft:1974bx}, have disclosed 
the appearance of non-renormalizable divergences already at one-loop in the presence 
of matter  \cite{'tHooft:1974bx,Deser:1974cz-1974zz}, and at two-loop for 
pure gravity \cite{Goroff:1985sz,vandeVen:1991gw}.
The situation is neither improved by the presence of a cosmological constant
 \cite{Christensen:1979iy}, nor by non-minimal 
couplings \cite{Gryzov-Barvinsky,Barvinsky:1993zg}.

The general conclusion usually taken out of these results is that general relativity is not fundamental and can only be quantized as an effective field theory. In this approach (see \cite{Donoghue:1994dn,Burgess:2003jk}), the gravitational action is organized in an energy expansion in curvature invariants. Once the scale for an experiment is identified, only the pertinent terms are then retained, allowing one to make predictions. A problem arises, however, once the energy is such that the curvature in Planck units reaches unity. At this point all curvature invariants are of the same order and an infinite number of couplings has to be fixed, so that the predictive power is lost.

A different conclusion can be attained if instead gravity turns out to be asymptotically safe (AS) \cite{Weinberg:1980gg} (see \cite{Niedermaier:2006wt,Percacci:2007sz,Reuter:2007rv,Litim:2008tt} for recent reviews). This scenario is based on Wilson's modern viewpoint
on renormalization \cite{Wilson:1973jj} and envisages the existence of a non-Gaussian fixed point (NGFP) of the renormalization group (RG) flow with a finite number of ultraviolet-attractive (relevant) directions. For RG trajectories attracted to the NGFP in the UV (spanning the UV critical surface of the fixed point), the fixed point ensures that the theory is free from uncontrollable UV-divergences, while the finite dimensionality of the surface ensures the predictivity of the theory at all energy scales. These criteria represent a non-perturbative analogue of the requirements underlying the usual perturbative renormalizability, which is recovered in the case of the fixed point being the Gaussian one.

In recent years, significant evidence for the asymptotic safety of gravity has been gathered by use of functional RG techniques \cite{ Reuter:1996cp,Souma:1999at,Lauscher:2001ya,Reuter:2001ag,Litim:2003vp, Granda:1998wn,Lauscher:2001rz,Lauscher:2001cq,Lauscher:2002mb, Codello:2007bd,Machado:2007ea,Codello:2008vh,us1,Forgacs:2002hz-Niedermaier:2002eq}, though support also comes from lattice simulations \cite{Ambjorn:2005db-2008wc}. The former approach generally employs a Functional Renormalization Group Equation (FRGE) for the effective average action originally derived in \cite{Wetterich:1992yh} and first applied to gravity in \cite{Reuter:1996cp}. Since an analysis based on the full equation is probably impossible, investigations usually rely on truncations of the theory space, whereby only a finite number of interaction-terms are retained. The reliability of the results found within such approximations can then be supported by considering their stability under a gradual extension of the truncation subspace. Indeed, all truncations studied so far,  from the Einstein-Hilbert-, to the $R^2$- and general $f(R)$-, up to the $R^2 + C^2$-truncations, give rise to a coherent picture, pointing at the existence of a NGFP dominating the UV behavior of gravity. 

A possible criticism on these results is that they are based on truncations which only contain interactions that are also unproblematic for the on-shell perturbative renormalizability. It is therefore a fundamental test for AS to include potentially dangerous terms in the truncation ansatz and study their effect on the fixed point structure of the theory. In pure gravity, the first non-trivial counterterm would be the Riemann-cube term of \cite{Goroff:1985sz,vandeVen:1991gw}. Including this term in the truncation ansatz is, however, technically very involved and beyond the current FRGE-techniques, even though the work presented in \cite{us1}, which for the first time permitted us to go beyond the class of $f(R)$-truncations, constitutes significant progress in that direction.

A technically less demanding, but equally illuminating, alternative is to study truncations for matter-coupled gravity. In this case the non-renormalizable counterterms already appear at one-loop and the occurrence of divergences proportional to $R^2$ and $C_{\m\n\rho\sigma}C^{\m\n\rho\sigma}$, which do not vanish on-shell, signal the break down of perturbative  renormalizability. To date, investigations of matter-coupled truncations, while also corroborating the asymptotic safety scenario, have remained restricted to the Einstein-Hilbert case \cite{Dou:1997fg,Granda:1997xk,Percacci:2002ie,Percacci:2003jz,Codello:2008vh}.  In the present paper, we go beyond this restriction, and study the non-perturbative RG flow of gravitational higher-derivative terms in the presence of a free, massless, minimally coupled scalar field, cf. eqs.\ \eqref{matter} and \eqref{HDflow} below. Anticipating our main result, we find that the NGFP previously reported for the Einstein-Hilbert case persists under the extension of the truncation subspace. This constitutes further evidence for the non-perturbative renormalizability of the theory, and, in particular, confirms that the non-renormalizable perturbative counterterms play no special role in the asymptotic safety scenario.

The rest of the paper is organized as follows. In Section~\ref{sec:pert} we review the counterterms arising from the perturbative quantization of general relativity coupled to a free scalar field, while in Section~\ref{sec:FRGE} we introduce the renormalization group methods employed. In Section~\ref{sec:EH} we revisit the Einstein-Hilbert truncation, for the pure gravity and matter-coupled cases, and finally in Section~\ref{sec:HD} we present the results for our full fourth-order truncation. We conclude with a discussion of our results in Section~\ref{sec:end}.
All the details of the calculations are cointained in the three appendices: Appendix~\ref{App:A} contains the Hessians entering the FRGE for our truncation ansatz, Appendix~\ref{App:B} presents the heat-kernel expansion for Lichnerowicz Laplacians, and finally Appendix~\ref{App:C} 
details the evaluation of the traces.

%----------------------------------------------------
\section{Perturbative non-renormalizability and counterterms}
\label{sec:pert}
%-------------------------------------------------------------
We start by reviewing the perturbative quantization of the Einstein-Hilbert action minimally coupled to a free scalar field. This provides the prototypical example of a gravitational theory which is perturbatively non-renormalizable at one-loop order \cite{'tHooft:1974bx}, as may be seen by computing its one-loop counterterms $\Delta\Gamma^{\rm div}$. In general, the one-loop effective action for a gauge theory takes the form
\be\label{1l}
\Gamma^{\rm 1-loop}[\Phi] = S[\Phi] + \frac{1}{2} \, {\rm STr} \ln \left[ \frac{\delta^2 S^{\rm tot}}{\delta \Phi^A \delta \Phi^B} \right] \, , 
\ee
where $\Phi^A$ is the full set of fields (including auxiliary fields and ghosts), $S^{\rm tot}[\Phi] = S[\Phi] + S^{\rm gf}[\Phi] + S^{\rm gh}[\Phi]$ is the total action of the theory including the gauge-fixing and ghost terms $S^{\rm gf}$ and $S^{\rm gh}$, and ${\rm STr}$ is a generalized functional trace carrying a minus sign for fermionic fields and a factor 2 for complex fields. Typically, this trace contains divergences which require regularization.

Our starting point is the action 
\be\label{action}
S[g, \phi] =  \int d^4x \sqrt{g} \, \left[ \kappa^{-2} (-R + 2 \Lambda) + \frac{1}{2} \, g^{\mu\nu} \, \p_\mu \phi \p_\nu \phi \right] \, ,
\ee
supplemented by the gauge-fixing term
\be\label{S:gf}
S^{\rm gf} = \frac{1}{2\kappa^2} \int d^4 x \sqrt{\gb} \, \gb^{\mu\nu} F_\mu F_\nu \, , \qquad F_\mu = \Db^\alpha h_{\mu \alpha} - \tfrac{1}{2} \Db_\mu h \, ,
\ee
and the corresponding ghost action. Here, $\kappa^2 = 16 \pi G$, $G$ and $\Lambda$ are the dimensionful Newton's and cosmological constant, respectively, $g_{\mu\nu}$ denotes the Euclidean space-time metric, and $\phi$ is a real scalar field. The gauge-fixing is carried out via the background field method, splitting the metric and scalar fluctuations into a background part, $\gb_{\mu\nu}, \bar{\phi},$ and fluctuations around this background, $h_{\mu\nu}, f$, according to $g_{\mu\nu} = \gb_{\mu\nu} + h_{\mu\nu}$ and $\phi = \bar{\phi} + f $.  Adapting the results \cite{Gryzov-Barvinsky,Barvinsky:1993zg} obtained via the Schwinger-DeWitt technique, the one-loop divergences arising from \eqref{action} are readily found to be\footnote{There is a typo in the coefficient of the squared potential in \cite{Barvinsky:1993zg}, the correct formula is given in \cite{Gryzov-Barvinsky}.}
\be\label{CT}
\begin{split}
\Delta\G^{\rm div}  = \frac{1}{(4 \pi)^2 \eps} \int d^4x \sqrt{g} & \, \Big[
\tfrac{43}{60} R_{\mu\nu} R^{\mu \nu} + \tfrac{1}{40} R^2 +\tfrac{213}{180} E + \tfrac{5}{4} \kappa^4  (\p_\mu \phi \p^\mu \phi)^2 \\ 
& - \kappa^2 (\tfrac{1}{3} R - 2 \Lambda) (\p_\mu \phi \p^\mu \phi) - \tfrac{26}{3} \Lambda R + 20 \Lambda^2
\Big] \, ,
\end{split}
\ee 
where $\eps = (d-4)$ and $E=C_{\m\n\r\s} C^{\m\n\r\s}-2 R_{\m\n}R^{\m\n}+ \tfrac{2}{3}R^2$ is the integrand of the Gauss-Bonnet term in four dimensions, with $C_{\m\n\r\s}$ being the Weyl tensor.

In order to get information on the renormalizability, the divergences \eqref{CT} have to be considered on-shell. 
The equations of motion resulting from \eqref{action} are
\be\label{eom}
\begin{split}
 \,D_\mu D^\mu \phi  = 0 \, , \quad
 \,R  = 4 \Lambda + \half \,\kappa^2 \,(\p_\mu \phi)^2 \, , \quad
 \,R_{\mu \nu}  = \Lambda g_{\mu\nu} + \half \,\kappa^2 \, \left[ \p_\mu \phi \, \p_\nu \phi \right] \, .
\end{split}
\ee
Substituting these, eq.\ \eqref{CT} can suggestively be written as\footnote{Note that this expression agrees both with the one-loop counterterm found by 't Hooft and Veltman \cite{'tHooft:1974bx} for $\L=E=0$, and with the one of Christensen and Duff \cite{Christensen:1979iy}
once the contribution of the scalar field is subtracted.}
\be\label{OSC}
\Delta\G^{\rm div} = \frac{1}{8 \pi^2 \eps} \int d^4x \sqrt{g} \left[ \frac{213}{360} E +\frac{203}{80} R^2-\frac{463}{20} R \Lambda +\frac{463}{10}\Lambda^2 \right] \, .
\ee
As the $R^2$ and $E$-terms are not of the form of the terms contained in the initial action, they cannot be absorbed by a renormalization of the coupling constants, indicating that the action \eqref{action} is indeed perturbatively non-renormalizable. The non-renormalizable on-shell counterterms are thus of fourth order in the gravitational sector and can be rewritten as
\be\label{NR}
\Delta \Gamma^{\rm NR} = \frac{1}{(4 \pi)^2 \eps} \int d^4x \sqrt{g} \left[ \frac{31}{18} \, R^2 + \frac{213}{180} \, C_{\mu\nu\rho \sigma} C^{\mu\nu\rho \sigma} \right] \, .
\ee

There is a common prejudice (see, for example \cite{Jacques}) that these interactions have a devastating effect also on the possible non-perturbative renormalizability (asymptotic safety) of the theory. Utilizing the new computational techniques developed in \cite{us1}, we will now show that this is not the case.

%-------------------------------------------------------------
\section{The functional renormalization group equation}
\label{sec:FRGE}
%-------------------------------------------------------------
A powerful tool in the study of the renormalization properties of a theory is the Functional Renormalization Group Equation (FRGE)  \cite{Wetterich:1992yh}
\be\label{FRGE}
\p_t \Gamma_k[\Phi, \bar{\Phi}] = \half {\rm STr} \left[ \left( \frac{\delta^{2} \Gamma_k}{\delta \Phi^A \delta \Phi^B } + \cR_k \right)^{-1} \, \p_t \cR_k  \right]\\,
\ee
where $\Phi$ denotes the physical fields and $\bar{\Phi}$ their background value.
The FRGE describes the dependence of the effective average action $\Gamma_k[\Phi, \bar{\Phi}]$ on the coarse-graining (or renormalization group) scale $k$. Here, $t = \log(k/k_0)$ and $\cR_k(p^2)$ is a (matrix-valued) infrared cutoff which provides a $k$-dependent mass-term for fluctuations with momenta $p^2<k^2$.  Apart from the requirement that it interpolates monotonically between $\cR_k (p^2) = 0$ as $p^2/k^2 \rightarrow \infty$ and $\cR_k (p^2) \propto k^2$ as $p^2/k^2 \rightarrow 0$, this cutoff can be arbitrarily chosen. For technical simplicity, our subsequent analysis will be based on the optimized cutoff \cite{Litim:2001up}, whose scalar part takes the form $R_k(p^2) = (k^2 - p^2) \theta(k^2 - p^2)$.

The FRGE has two key features, owing mainly to the IR regulator structure. First, its solutions interpolate between the ordinary effective action $\Gamma \equiv \Gamma_{k \rightarrow 0}$ and an initial action $\Gamma_\Lambda$ at the UV cutoff scale, 
which  in the limit $\Lambda \rightarrow \infty$ essentially reduces to the bare action (see \cite{Manrique:2008zw} for more details). The effective average action is  obtained by integrating out modes in the path integral from a UV cutoff scale $\L$ down to the scale $k$, as in a Wilsonian coarse graining procedure, with the modes below $k$ being suppressed. Secondly, due to the derivative $\p_t \cR_k(p^2)$ in the numerator, the contributions to the flow equation are localized on modes with momenta near $k^2$, so that the trace remains finite and locally well-defined at all scales. In particular, while a theory might require a UV regulator at the level of its path integral, at the FRGE level this UV regularization is superfluous. 

The main shortcoming of the FRGE, however, is that it cannot be solved exactly.
In order to extract physics from it, one therefore has to resort to approximations.
One possibility is, of course, perturbation theory. In the one-loop approximation, where $\Gamma_k$ under the STr is replaced by the $k$-independent bare action, 
 one then recovers upon integration the usual non-renormalizable logarithmic divergences \cite{Codello:2008vh}.

Going beyond perturbation theory, a standard approximation scheme is the truncation of the RG flow, whereby the flow of the full theory is projected onto a subspace spanned by a finite number of interaction monomials. Making an ansatz $\Gamma_k[\Phi, \bar{\Phi}] = \sum u_i(k) \, \cO_i[\Phi, \bar{\Phi}]$ with a finite subset of the interaction monomials  $\cO_i$ and substituting this ansatz into the FRGE, this technique allows one to extract the $\beta$-functions for the dimensionful coupling constants $u_i$. 
When analyzing the properties of the RG flow, it is then most convenient to switch to the dimensionless coupling constants $g_i = k^{-d_i} u_i$, with $d_i$ being the mass-dimension of $u_i$, which results in autonomous $\beta$-functions $\p_t g_i = \beta_{g_i}(g_i)$.
Within Wilson's modern perspective on renormalization, the renormalizability of the theory is then determined by the fixed points (FP) $\{g_i^*\}$ of the 
$\b$-functions, $\{\b_{g_i}(g_i^*)=0\}$. Around any such FP, the linearized RG flow $\p_t g_i = {\bf B}_{ij} (g_j - g^*_j)$ is governed by the stability matrix 
\be
{\bf B}_{ij} = \left. \p_j \beta_i \right|_{\{g_i^*\}}\ .
\ee
Defining the stability coefficients $\th_i$ as minus the eigenvalues of $\bf B$, the relevant (irrelevant) directions are associated to the eigenvectors
corresponding to stability coefficients with a positive (negative) real part.

In general, it is useful to cast the effective average action into the form \cite{Reuter:1996cp}
\be
\Gamma_k[\Phi, \bar{\Phi}] = \bar{\Gamma}_k[\Phi] + \widehat{\Gamma}_k[\Phi - \bar{\Phi}, \bar{\Phi}] + 
S^{\rm gf}[g, \gb] + S^{\rm gh}[g, \gb, {\rm ghosts}] \, .
\ee
 In this decomposition $\bar{\Gamma}_k[\Phi]$ depends on the physical fields only, and $S^{\rm gf}$ and $S^{\rm gh}$ denote the {\it classical} gauge-fixing and ghost-terms respectively. $\widehat{\Gamma}_k$ encodes the deviations
$\Phi - \bar{\Phi}$, thus vanishing for $\Phi = \bar{\Phi}$, and captures the quantum corrections to the gauge-fixing and ghost sector of the effective average action.
For the remainder of this work, we will focus on truncations of the form
\be\label{EAA}
\bar{\Gamma}_k[\Phi] = \Gamma_k^{\rm grav}[g] + \Gamma^{\rm matter}[g, \phi] \, .
\ee
Here $\Gamma_k^{\rm grav}$ is the gravitational part of the effective average action, for which we will specify two different truncations in Sec.~\ref{sec:EH} and Sec.~\ref{sec:HD} while
\be\label{matter}
\Gamma^{\rm matter}[g, \phi] = \half \, \int d^4x \sqrt{g} \,  \, g^{\mu\nu} \, \p_\mu \phi \, \p_\nu \phi
\ee
is the ($k$-independent) action for a minimally coupled free scalar field. Furthermore, we set $\widehat{\Gamma}_k[\Phi-\bar{\Phi}, \bar{\Phi}] = 0$ in the sequel.\footnote{Of course, it would also be desireable to obtain a better understanding of the influence of $\widehat{\Gamma}_k[\Phi-\bar{\Phi}, \bar{\Phi}]$ on the RG flow. 
 In this context, the adaptation of the background independent version of $\Gamma_k$, discussed for Yang-Mills theories in \cite{Litim:2002ce}, could provide a valuable tool.} For $S^{\rm gf}$, we consider the following generalization of \eqref{S:gf}, 
which allows for a straightforward application to gravitational actions including higher-derivative terms
\be\label{S:gf2}
S^{\rm gf} = \frac{1}{2} \int d^4 x \sqrt{\gb} \, F_\mu \, Y^{\mu\nu} F_\nu \, , \quad
F_\mu = \Db^\nu h_{\mu \nu} - \tfrac{1+\rho}{4} \Db_\mu h \, , \quad Y^{\mu \nu} = \big[ \frac{\alpha}{\kappa^2} + \beta \Db^2 \big] \gb^{\mu \nu} \, .
\ee
The gauge-fixing \eqref{S:gf} is obtained as the limit $\rho = 1,\ \alpha = 1,\ \b=0$. When  analyzing the RG flows in sections \ref{sec:EH} and \ref{sec:HD}, however, it will be more convenient to set 
\bea
& \rho = 0,\ \alpha \rightarrow \infty,\ \beta = 0, \ & \,\,\, \text{for the Einstein-Hilbert truncation}, \label{gfEH} \\
& \rho = 0,\ \alpha = 0,\ \beta \rightarrow \infty, \ & \,\,\, \text{for the higher-derivative truncation}. \label{gfHD}
\eea
Here, the arrow indicates that the limit is to be taken under the trace of the flow equation.

The key step in utilizing \eqref{FRGE} for extracting $\beta$-functions
is the evaluation of the operator trace appearing on its r.h.s. For the ansatz \eqref{EAA}, this STr decomposes into a trace over the gravitational and the matter sector, respectively. 

The contribution from the gravitational sector is obtained as follows \cite{us1} (for further details, see the Appendices). We first compute the
second variation of $\Gamma^{\rm grav}_k$ and perform a transverse-traceless decomposition of the metric fluctuations and the ghost fields, dealing with the Jacobians arising from this decomposition by introducing suitable auxiliary fields as in the Faddeev-Popov trick. The subsequent computations can then be simplified by identifying $g$ with a suitable class of background metrics $\gb$. This class must be general enough to distinguish 
the interaction monomials contained in $\Gamma^{\rm grav}_k$ and, at the same time, simple enough to ease the evaluation of the operator traces.
For our present purposes, it suffices to consider the class of generic compact Einstein backgrounds (without Killing or conformal Killing vectors), not necessarily solving the equations of motion \eqref{eom}. Utilizing these backgrounds, all differential operators within our particular traces organize themselves into Lichnerowicz form 
\be\label{def:LL}
\begin{split}
\Delta_{2L} \phi_{\mu\nu} & \equiv -D^2 \phi_{\mu\nu} - 2 R_{\mu\,\,\,\nu}^{\,\,\,\alpha\,\,\,\beta} \phi_{\alpha\beta}\,,\quad
\Delta_{1L} \phi_\mu  \equiv \left[ -D^2  - \tfrac{1}{4}  R \right] \,  \phi_\mu\,,\quad
\Delta_{0L} \phi  \equiv -D^2 \phi\, ,
\end{split}
\ee
i.e., minimal second order differential operators $\D_{sL} = - \Db^2 + {\bf Q}_{s}$, with spin-dependent matrix potentials ${\bf Q}_s$ acting on transverse-traceless matrices ($s=2$), transverse vectors ($s=1$) and scalars ($s=0$). This feature is crucial for the non-perturbative evaluation of the traces, as it makes them amenable to standard heat kernel techniques without having to resort to non-minimal (or $k$-dependent) differential operators. 
The final steps in this computation then follow the standard FRGE procedure (see e.g.\ \cite{Reuter:2007rv}). 
First, the cutoff operators $\cR_{s,k}$ are constructed in such a way that the modified propagators are obtained by replacing $\D_{sL}\to P_{s,k}(\D_{sL})=\D_{sL}+\cR_{s,k}(\D_{sL})$. Then, the traces in the flow equation are evaluated using the ``early-time expansion" of the heat kernel adapted to the Lichnerowicz Laplacians. 

Written in terms of operator traces, the flow equation then takes the following generic form
\be\label{Ftrace}
\p_t \Gamma_k[\Phi, \bar{\Phi}] = \cS_{\rm 2T} + \cS_{\rm hh} + \cS_{\rm 1T} + \cS_{\rm 0} + n_s \, \cS_{\rm matter} \, , 
\ee
where $n_s$ gives the number of matter fields and the subscripts ``2T'', ``1T'' and ``0'' indicate traces taken on the space of symmetric transverse-traceless matrices, transverse vectors and scalars, respectively. Applying the background-field method (setting $\bar{\phi} = 0$, for convenience),
 it is straightforward to find the matter contribution
to the flow equation \cite{Dou:1997fg},
\be\label{mattr}
\cS_{\rm matter} = \frac{1}{2} \Tr_0 \left[ \frac{\p_t R_{0,k}}{P_{0,k}} \right] \, .
\ee
Furthermore, owed to the special gauge choices \eqref{S:gf2} with \eqref{gfEH} and \eqref{gfHD},
$\cS_{\rm 1T}$ and $\cS_{\rm 0}$ are universal, in the sense that they are independent of
the particular $\Gamma_k^{\rm grav}[g]$ adopted here,
\be\label{Tr:ghost}
\cS_{\rm 1T} = \, - \frac{1}{2} \Tr_{\rm 1T} \left[ 
\frac{\p_t R_{1,k}}{P_{1,k}}
\right] \, , \qquad  
\cS_{\rm 0} = \, - \frac{3}{2} \Tr_{\rm 0} \left[ 
\frac{ \p_t R_{0,k}}{3 P_{0,k} - R}
\right] \, .
\ee
Following the computations outlined in Appendix \ref{App:C}, the evaluation of the traces can be carried out using 
the early-time heat-kernel expansion for Lichnerowicz-operators. The resulting expressions are given in eqs.\ \eqref{mattercont} and \eqref{eq:TRuni}, respectively. We should also note that it is straightforward to consider the effect of a number $n_s$ of scalar fields by adding further copies of \eqref{mattercont} to the RG equations. In what follows, however, we will mostly focus on the cases $n_s =0,1$, keeping the label $n_s$ only to highlight the matter contribution.

%------------------------------------------------------------------------------
\noindent
\section{Fixed points of the Einstein-Hilbert truncation}
\label{sec:EH}
%------------------------------------------------------------------------------
In this section, we approximate $\Gamma_k^{\rm grav}$ by the Einstein-Hilbert action
with scale dependent coupling constants,
\be\label{EHtrunc}
\Gamma_k^{\rm grav}[g] = \frac{1}{16 \pi G_k} \int d^4x \sqrt{g} \left( -R + 2 \Lambda_k \right) \, .
\ee
This truncation has already been considered in the context of pure gravity in \cite{Reuter:2001ag,Litim:2003vp,Souma:1999at,Lauscher:2001ya} and in the matter-coupled case in \cite{Dou:1997fg,Percacci:2002ie,Percacci:2003jz,Codello:2008vh}.
Here, we complement this analysis by implementing the gauge-fixing (\ref{S:gf2}-\ref{gfEH}) and organizing the operators inside the trace in terms of Lichnerowicz Laplacians on a generic Einstein space, lending further evidence to the robustness of these earlier works.

Using the results of Appendix \ref{App:A}, the non-universal traces resulting from \eqref{EHtrunc}
are 
\be\label{NU-EH}
\cS_{\rm 2T} = \frac{1}{2} \Tr_{\rm 2T} \left[ \frac{\p_t(u_1 R_k)}{u_1 P_{2,k} + u_1 R/2 + u_0} \right] \, , 
\quad
\cS_{hh} = \frac{3}{2} \Tr_{\rm 0} \left[ \frac{\p_t(u_1 R_k)}{3 u_1 P_{0,k} + 2 u_0} \right] \, ,
\ee
where the coupling constants $u_i$ are defined in \eqref{couplings}.
Adapting the computations outlined in \cite{Reuter:2007rv,Codello:2008vh} to the steps described in Section \ref{sec:FRGE}, the $\beta$-functions of the dimensionless Newton's constant $g_k = G_k \, k^2$ and cosmological constant $\lambda_k =  \Lambda_k \, k^{-2}$ then
read
\be \label{betaEH}
\beta_g = (2 + \eta_N^{\rm EH}) g \, , \quad \beta_\lambda = (\eta_N^{\rm EH}-2) \lambda + \frac{g}{2 \pi} B_3(\lambda) - \frac{g}{4\pi} \eta_N^{\rm EH} 
\left(5 \Pt^1_2(-2\lambda) + \Pt^1_2(-\tfrac{4}{3}\lambda) \right)\, ,
\ee
where,
\be\label{etaEH}
\eta_N^{\rm EH} = \frac{g B_1(\lambda)}{6 \pi - g B_2(\lambda)}
\ee 
denotes the anomalous dimension of Newton's constant and where we have defined
\be\label{BEH}
\begin{split}
B_1(\lambda) = & \, (n_s - 7) \Phi^1_1(0) - 2 \Phi^2_2(0) + \Phi^1_1(-\tfrac{4}{3}\lambda)  - 10 \Phi^1_1(-2\lambda) - 15 \Phi^2_2(-2\lambda) \, , \\
B_2(\lambda) = & \, 5 \Pt^1_1(-2\lambda) + \tfrac{15}{2} \Pt^2_2(-2\lambda) - \tfrac{1}{2} \Pt^1_1(- \tfrac{4}{3} \lambda) \, , \\
B_3(\lambda) = & \, (n_s - 4) \Phi^1_2(0) + \Phi^1_2(-\tfrac{4}{3}\lambda) + 5 \Phi^1_2(-2\lambda) \, . 
\end{split}
\ee
We can see that the inclusion of matter fields simply results in a shift by a constant in the equations above. In a sense, ``small'' values of $n_s$ could therefore be interpreted as a ``perturbation'' of the $\beta$-functions for pure gravity, which are recovered in the limit $n_s = 0$.
Note that these $\beta$-functions are non-perturbative, in the sense that the anomalous dimension $\eta_N^{\rm EH}$ contains infinitely many powers of $g$. 
 
Analyzing the fixed point structure of the $\beta$-functions \eqref{betaEH}, we first note that both the one-loop and the non-perturbative $\beta$-functions possess a GFP at $\lambda^* = 0,\ g^* = 0$ for all values of $n_s$, corresponding to the free theory. Its stability coefficients are given by the canonical mass-dimensions of $\Lambda$ and $G$, that is, $\theta_1 = 2$ and $\theta_2 = -2$. Owing to the negative mass dimension of Newton's constant, the only relevant direction is at $G=0$, amounting to a trivial theory with just the cosmological term. As soon as we turn on Newton's constant, the flow is carried away from the GFP in the UV. Thus, our gravity-matter theory is not inside the UV critical surface of the GFP, verifying its perturbative non-renormalizability from the Wilsonian viewpoint. Note that this behavior is independent of $n_s$.

Remarkably, the $\beta$-functions \eqref{betaEH} also give rise to a NGFP at positive values $\l^*>0, g^* > 0$. Its position and stability coefficients for $n_s=0,1$ are given in Tables \ref{t.FPEH} and \ref{t.FPEHs} (together with a comparison to earlier works). Note that this NGFP is UV-attractive for both the dimensionless Newton's constant and the cosmological constant. The gravity-matter theory considered here is within the UV critical surface of the NGFP. In other words, matter-coupled gravity in the Einstein-Hilbert truncation is asymptotically safe. Note also that making the transition from pure gravity to gravity coupled to a free scalar field has a rather small effect on the numerical values obtained for the NGFP.  

\begin{table}[t!]
\begin{center}
\begin{tabular}{|c|cc|c|c||cc|c|}\hline
Ref.\ & \; \; \; $g^*$ \; \; \; & \; \; \; $\lambda^*$ \; \; \; & \; \; \; $g^* \lambda^*$ \; \; \; & \; \; \; $\theta^\prime \pm \I \theta^{\prime \prime}$ \; \; \; & \; \; \; $\alpha$ \; \; \; & \; \; \; $\rho$ \; \; \; & \; \; \; cutoff \; \; \; \\ \hline
Here & $0.902$ & $0.109$ & $0.099$ & $2.52 \pm 1.78 \I$ & $\infty$ & $0$ & II, opt 
\\ \hline
RS & $0.403$ & $0.330$ & $0.133$ & $1.94 \pm 3.15\I$ & $1$ & $1$ & I, sharp
\\ \hline
LR & $0.272$ & $0.348$ & $0.095$ & $1.55 \pm 3.84\I$ & $1$ & $1$ & I, exp \\
& $0.344$ & $0.339$ & $0.117$ & $1.86 \pm 4.08\I$ & $\infty$ & $1$ & I, exp  
\\ \hline
L & $3 \pi/8$ & $0.25$ & $3 \pi/32$ & $1.67 \pm 4.31 \I $ & $\infty$ & $1$ & I, opt\\ \hline
CPR & $0.707$ & $0.193$ & $0.137$ & $1.48 \pm 3.04\I$ & $1$ & $1$ & I, opt \\ 
& $0.556$ & $0.092$ & $0.051$ & $2.43 \pm 1.27 \I$ & $1$ &$1$ & II, opt \\
& $0.332$ & $0.274$ & $0.091$ & $1.75 \pm 2.07\I$ & $1$ & $1$ & III, opt \\ \hline
\end{tabular}
\parbox[c]{\textwidth}{\caption{\label{t.FPEH}{Einstein-Hilbert truncation: comparison of the characteristics of the pure gravity NGFP obtained here employing the ``universal gauge-fixing'', and results reported earlier in the literature \cite{Reuter:2001ag} (RS), \cite{Lauscher:2001ya} (LR), \cite{Litim:2003vp} (L), and \cite{Codello:2008vh} (CPR), respectively. The cutoff classification as Type I, II or III follows \cite{Codello:2008vh}, while with ``opt", ``sharp" and ``exp" we refer to the shape function being of the optimized, sharp or exponential type respectively. The fixed point is robust under variation of the gauge-fixing parameter $\alpha$, the shape function used in the IR regulator, and the implementation of the regularization.}}}
\end{center}
\end{table}

\begin{table}[t!]
\begin{center}
\begin{tabular}{|c|cc|c|c||cc|c|}\hline
Ref.\ & \; \; \; $g^*$ \; \; \; & \; \; \; $\lambda^*$ \; \; \; & \; \; \; $g^* \lambda^*$ \; \; \; & \; \; \; $\theta^\prime \pm \I \theta^{\prime \prime}$ \; \; \; & \; \; \; $\alpha$ \; \; \; & \; \; \; $\rho$ \; \; \; & \; \; \; cutoff \; \; \; \\ \hline
Here & $0.860$ & $0.131$ & $0.112$ & $2.58 \pm 1.95 \I$ & $\infty$ & $0$ & II, opt 
\\ \hline
PP & $0.254$ & $0.366$ & $0.093$ & $1.71 \pm 4.16\I$ & $1$ & $1$ & I, exp \\
& $0.320$ & $0.359$ & $0.115$ & $2.08 \pm 4.38\I$ & $\infty$ & $1$ & I, exp  
\\ \hline
\end{tabular}
\parbox[c]{\textwidth}{\caption{\label{t.FPEHs}{Einstein-Hilbert truncation: comparison of the characteristics of the NGFP obtained from gravity coupled minimally to a real scalar field. The first line is obtained from the ``universal gaugefixing'', while the data of the second and third line has been obtained in \cite{Percacci:2002ie,Percacci:2003jz} (PP), and is given for further comparative purposes. Again, the fixed point is robust under variation of the gauge-fixing parameter $\alpha$, the shape function used in the IR regulator, and the implementation of the regularization.}}}
\end{center}
\end{table}

%------------------------------------------------------------------------------
\section{Fixed points of the higher-derivative-matter truncation}
\label{sec:HD}
%------------------------------------------------------------------------------
The key question raised by the results obtained within the Einstein-Hilbert truncation is whether the resulting fixed points survive in the full theory and, in particular, whether the NGFP will persist, with similar characteristics, once the perturbative counterterms 
\eqref{NR} are included in the truncation subspace. While the former is a million-dollar question (recession notwithstanding), the latter can be answered positively. 

To this effect, we enhance the truncation subspace \eqref{EHtrunc} and consider the ansatz 
\be\label{HDflow}
\Gamma_k^{\rm grav}[g] = \int d^4x \sqrt{g} \left[ \frac{1}{16 \pi G_k} (-R + 2 \Lambda_k) - \frac{\omega_k}{3 \sigma_k} R^2 + \frac{1}{2 \sigma_k} C_{\mu\nu\rho\sigma} C^{\mu\nu\rho\sigma}   + \frac{\theta_k}{\s_k} E\right] \, ,
\ee
which is precisely of the form Einstein-Hilbert-action plus perturbative counterterms. In the spirit of the RG, the numerical coefficients in the latter have been replaced by the canonical (scale-dependent) coupling constants. 

Following the derivation given in Appendix~\ref{App:A}, 
the gravitational contribution to $\eqref{Ftrace}$
enters into $\cS_{\rm 2T}$ and $\cS_{hh}$ only and is given by
\be\label{Tr:grav}
\begin{split}
\cS_{\rm 2T} = & \,  \frac{1}{2} \Tr_{\rm 2T} 
\left[
\frac{\p_t \left\{ 2 u_3 (P_{2,k}^2 - \D_{2L}^2)  - (u_1 + u_\flat R) R_{2,k} \right\}}
{2 u_3 P_{2,k}^2 - (u_1 + u_\flat R ) P_{2,k} - \tfrac{1}{2} u_1 R - u_0} 
\right] \, , \\
\cS_{hh} = & \, \frac{1}{2} \Tr_{\rm 0} \left[
\frac{\p_t \left\{ 6 u_2 (P_{0,k}^2 - \D_{0L}^2) + (u_1 - 2  u_2 R) R_{0,k}  \right\}}{6 u_2 P_{0,k}^2 + (u_1 - 2 u_2 R) P_{0,k} + \tfrac{2}{3} u_0}
\right] \, . \\
\end{split}
\ee
 The coupling constants appearing in these expressions are related to \eqref{HDflow} via 
 \be\label{couplings}
 u_0 = \frac{\Lambda_k}{8 \pi G_k},\quad
 u_1 = -\frac{1}{16 \pi G_k},\quad
 u_2 = - \frac{\omega_k}{3 \sigma_k}+ \frac{\theta_k}{6\s_k},\quad
 u_3 = \frac{1}{2 \sigma_k}+\frac{\theta_k}{\s_k},
 \ee
 and $u_\flat = 2u_2-\tfrac{1}{3}u_3$. Note that, because of the Einstein-space choice, we can distinguish only two of the three higher-derivative couplings.
Lastly, note also that including $\cS_{\rm matter}$ has a similar effect as in the Einstein-Hilbert case, leading to shifts in certain coefficients appearing in the $\beta$-functions.

The projection of the traces onto the truncation subspace spanned by \eqref{HDflow}
can again be carried out utilizing the early time expansion of the heat kernel adapted to the Lichnerowicz operators on a general Einstein background, as detailed in Appendix~\ref{App:B}. Following the computation outlined in Appendix~\ref{App:C}, and introducing the dimensionless couplings
\be\label{dimless}
g_0 = u_0 k^{-4} \; , \quad g_1 = u_1 k^{-2} \; , \quad g_2 = u_2 \, , \quad g_3 = u_3 \, ,
\ee
the $\beta$-functions following from our truncation
are given by
\be\label{beta}
\begin{split}
\p_t g_0 = & \, -4 \, g_0 + \frac{1}{2(4\pi)^2} \,  \left\{ C_1 + \tC_1 +(2 n_s- 8) \Phi^1_2 \right\} \, , \\
 \p_t g_1 = & \, -2 \, g_1 + \frac{1}{2(4\pi)^2} \, \left\{ C_2 + \tC_2 + \tfrac{n_s-7}{3} \Phi^1_1 - \tfrac{2}{3} \Phi^2_2 \right\} \, , \\
 \p_t g_3 = & \, \frac{1}{(4 \pi)^2} \left\{ \tfrac{1}{360} C_3 + \tfrac{5}{9} \tC_3 + \tfrac{11+n_s}{360} \varphi \right\} \, , \\
 \p_t g_2 - \tfrac{1}{6} \p_t g_3 = & \,  \frac{1}{(4 \pi)^2} \left\{
 \half C_4 + \half \tC_4 - \tfrac{1}{18} \Phi^2_1 - \tfrac{1}{9} \Phi^3_2 +(\tfrac{n_s}{160}- \tfrac{1}{15}) \varphi 
 \right\} \, ,  
\end{split}
\ee
where the threshold functions $\Phi_{n}^{p}$,  $\varphi$ are respectively defined in \eqref{defPhi} and \eqref{defPhiOpt}, with the former evaluated at zero argument.
The expansion coefficients $C_i$ and $\tC_i$ arise from evaluating the traces $\cS_{\rm 2T}$ and $\cS_{hh}$, respectively, and are defined in eqs.\ \eqref{Ccoeff} and \eqref{tCcoeff}. For notational reasons, the $\beta$-functions \eqref{beta} are given implicit form. In particular, we stress that {\it both} the left and right-and-side contains derivatives $\p_t g_i$. The ``standard''  $\beta$-functions $\p_t g_i = \beta_i(g_i)$ can then be obtained by solving these equations for $\p_t g_i$, which
can be straightforwardly done using algebraic manipulation software. 

The resulting expressions may again be expanded for small $g$ and, here, $\sigma$. In this respect, we first note that \eqref{beta} contains contributions from arbitrary powers in $g, \sigma$, and hence that the $\beta$-functions capture some truly non-perturbative information. Secondly, we have verified that the leading contributions in this expansion reproduce the known universal parts of the one-loop $\beta$-functions in higher-derivative gravity, providing an important confirmation of the correctness of our derivation.

Remarkably, the fixed point structure originating from these higher-derivative $\beta$-functions is very similar to the Einstein-Hilbert case. First, we recover the two generalizations of the GFP, familiar from perturbation
theory\footnote{In a slight abuse of notation, we neglect the topological term here, setting $\theta = 0, \p_t \theta= 0$. Also note that the existence of this GFP is actually compatible with the analysis \cite{Lauscher:2002mb}, which did not consider the case of the inverse coupling $1/\sigma \rightarrow \infty$.},
\be
g^* = 0  \, , \qquad \lambda^* = 0 \, , \qquad \sigma^* = 0 \, , \qquad 
\omega^*_{1,2} = - \tfrac{1}{120} \left(90 \pm \sqrt{15708 - 101 n_s} \right) \, ,
\ee
existing for $0 \le n_s \le 155$ and with
stability properties given by the following eigensystem
\be
\begin{array}{llll}
\theta_1 = 2 \, , \qquad & V_1 =  \{ 1,0,0,0 \}^{\rm T} \, , \qquad \quad &
\theta_2 = -2 \, , \qquad & V_2 =  \{ \frac{2+n_s}{16 \pi},1,0,0 \}^{\rm T} \, , \\
\theta_3 = 0 \, , \qquad & V_3 =  \{ 0,0,1,0 \}^{\rm T} \, , \qquad \quad &
\theta_4 = 0 \, , \qquad & V_4 = \{ 0,0,0,0 \}^{\rm T} \, . 
\end{array}
\ee
These GFPs correspond to the free theory,  and their stability coefficients are given by the canonical mass dimension of the corresponding (dimensionful) couplings. In particular, the eigendirection associated with Newton's constant is still UV repulsive, while the directions associated with the new couplings $\sigma_k, \omega_k$ are marginal. Going beyond the linear approximation, the
marginal directions are found to be UV-attractive, in accordance with the one-loop calculations
\cite{Codello:2006in}.

Most importantly, the matter-coupled higher-derivative truncation also gives rise to the generalization of the NGFP. Its corresponding position and stability coefficients are given in Tables \ref{t.FP} and \ref{t.Stab} (under the entries ``$R^2+C^2+$scalar''). 
For completeness, these tables also include the data on the NGFP for the pure gravity case (``$R^2+C^2$"), first reported in \cite{us1}, and we note that its properties are again very similar to those in the gravity-matter case, thus giving rise to essentially the same picture. 

One salient difference with the Einstein-Hilbert case is the fact that all stability coefficients are now real. This is in agreement with the one-loop results of \cite{Codello:2006in}, but it is surprising that, unlike in the Einstein-Hilbert case, the transition from the one-loop to the non-perturbative treatment does not give rise to complex eigenvalues. We can trace this result to the contribution of the $C^2$ terms coming out of the traces: indeed, restricting our computation to a spherically symmetric space we again find complex eigenvalues.

Crucially, increasing the dimension of the truncation subspace, with respect to the Einstein-Hilbert case, adds one UV-attractive and one UV-repulsive eigendirection to the stability matrix, so that the UV critical hypersurface in the truncation subspace is now three-dimensional. We then have a three-dimensional subspace of RG trajectories which are attracted to the NGFP in the UV and are therefore ``asymptotically safe''.  Thus, non-perturbative renormalizability persists also in the presence of the 
one-loop perturbative counterterms in the truncation ansatz.

\begin{table}[t!]
\begin{center}
\begin{tabular}{|c||c|c|c|c|c||c|c|}\hline
Truncation & \; \, $g^*$ \; \, & \; \, $\lambda^*$ \; \, & \; \, $u_2$ \, \; & \; \, $u_3$ \, \; & \; \,$u_4$ \, \;& \; \, $\lambda^* g^*$ \, \; & cutoff \\ \hline
$R^2 + C^2$ & $1.960$ & $0.218$ & $0.008$ & $-0.0050$ & -- &   $0.427$ & II, opt \\
LR II  & $0.292$ & $0.330$ & $0.005$ & -- & -- &  $0.096$ & I, exp \\
CP &  $1.389$  &  $ 0.221 $  & $*$ & $*$ & $*$ & $ 0.307$   & \; perturbative one-loop  \; \\ \hline \hline
\; $R^2 + C^2 + $ scalar \; & 
$2.279$ & $0.251$ & $0.010$ & $-0.0043$ & -- &   $0.571$ & II, opt \\ \hline
\end{tabular}
\parbox[c]{\textwidth}{\caption{\label{t.FP}{Position of the NGFP optained from the {\it non-perturbative} $\beta$-functions of the $R^2 + C^2$-truncation, eq.\ \eqref{beta}. For comparison, we also give the data of the $R^2$-truncation \cite{Lauscher:2002mb} (LR II), and the perturbative one-loop result \cite{Codello:2006in} (CP). In the latter line, the $*$ indicates that $\omega^* =-0.0228, \theta^* = 0.327$ approach finite values in the UV, while $\sigma$ runs logarithmically to zero, realizing the asymptotic freedom of the one-loop result. The last line gives the position of the NGFP upon including a minimally coupled scalar field.}}}
\end{center}
\end{table}

\begin{table}[t!]
\begin{center}
\begin{tabular}{|c||c|c|c|c|} \hline
Truncation & $\theta_1$ & $\theta_2$ & $\theta_3$ & $\theta_4$  \\ \hline
$R^2 + C^2$ &  $2.51$ & $1.69$ & $8.40$ & \; \;  $-2.11$ \; \;  \\
LR II & \; \; $2.15 + 3.79 \I$ \; \; & \; \; $2.15 - 3.79 \I$ \; \;& \; \; $28.8$ \; \; & \; \; -- \; \;  \\
CP &  $4$ &$2$ & $*$ & $*$      \\ \hline \hline
\; $R^2 + C^2 + $ scalar \; & $2.67$ & $1.39$ & $7.86$ & $-1.50$ \\ \hline
\end{tabular}
\parbox[c]{\textwidth}{\caption{\label{t.Stab}{Stability coefficients of the NGFP optained from the {\it non-perturbative} $\beta$-functions of the $R^2 + C^2$-truncation, eq.\ \eqref{beta}. For comparison, we also give the data of the $R^2$-truncation \cite{Lauscher:2002mb} (LR II), and the perturbative one-loop result \cite{Codello:2006in} (CP). In the latter line, the $*$ indicates the logarithmic running of the marginal coupling constants towards asymptotic freedom. The last line gives the stability coefficients of the NGFP upon including a minimally coupled scalar field.}}}
\end{center}
\end{table}

%------------------------------------------------------------------------------
\section{Discussion and conclusion}
\label{sec:end}
%------------------------------------------------------------------------------
In this paper, we have analyzed the fixed point structure underlying the renormalization group (RG) flow of gravity minimally coupled to a free scalar field, within a truncation approximation. From the viewpoint of perturbative quantization, this setup provides a prototypical example of a quantum theory of gravity which is perturbatively non-renormalizable at the one-loop level \cite{'tHooft:1974bx}. Here, higher-derivative interactions arise as perturbative counter\-terms, signaling the presence of divergences which cannot be absorbed by the renormalization of the coupling constants.  However,
despite the breakdown of the perturbative quantization scheme, there is the possibility that this gravity-scalar theory constitutes a well-defined and predictive quantum theory within the realm of asymptotic safety \cite{Percacci:2002ie,Percacci:2003jz}. With this in mind, we first considered the case of the Einstein-Hilbert truncation, before extending it to a higher-derivative truncation by including the interactions of the form of the one-loop counterterms. 

As our main result, we show that all these truncations give rise to a non-Gaussian fixed point, which underlies the conjectured asymptotic safety of the theory, in addition to a Gaussian fixed point linked to the perturbative quantization. Both fixed points are robust under the extension of the truncation subspace by higher derivative terms. This result explicitly shows that, contrary to a common worry, the inclusion of perturbative counterterms in the truncation subspace of a gravity-matter theory has {\it no qualitative effect} on its fixed point structure. In particular, we find {\it no indication} that these interactions are fatal to non-perturbative renormalizability of the theory. 

A second remarkable property surfaces when comparing the fixed point structure obtained for pure gravity ($n_s = 0$) and gravity coupled to one free scalar ($n_s = 1$) given in the top and bottom lines of Tables \ref{t.FP} and \ref{t.Stab}, respectively. Including the scalar field shifts the fixed point values obtained for pure gravity only very mildly, so that the resulting fixed point patterns are very similar. In a sense, this indicates that (at least for the present truncations) the UV behavior of the gravity-matter theory is still dominated by its gravitational sector, so that it still behaves ``essentially gravitational'' at high energies. 
Following \cite{Percacci:2002ie,Gies:2009hq}, it would be very interesting to determine which matter sectors lead to asymptotically safety 
gravity-matter theories (which we might dub the ``asymptotic safety territories''), 
taking the higher-derivative terms \eqref{HDflow} into account.

While our results on the interplay between the perturbative counterterms and asymptotic safety in the gravity-matter case are already trend-setting, it would nevertheless be desirable to carry out an analogous computation for pure gravity, where non-renormalizable divergences set in at two-loop level \cite{Goroff:1985sz,vandeVen:1991gw}. This is, however, still beyond the current technical scope of the  functional renormalization group techniques employed in this paper. Nevertheless, various arguments have been put forward \cite{Codello:2008vh,Reuter:2008qx} that the situation there will be similar to the one encountered here: perturbative counterterms are likely to have no special effect on the asymptotic safety of the theory. \\[3ex]

\noindent
{\it Acknowledgments} -- 
We thank J.\ Distler and R.\ Percacci for animated discussions which triggered this investigation. Furthermore, we are greatful to J.P.\ Blaizot, A.\ Codello, R.\ Loll, E.\ Manrique, and M.\ Reuter for useful conversations. Research at Perimeter Institute is supported in part by the Government of Canada through NSERC and by the Province of Ontario through MRI. P.F.M. is supported by the Netherlands Organization for Scientific Research (NWO) under their VICI program. F.S.\ acknowledges 
financial support from the ANR grant BLAN06-3-137168.

\begin{appendix}
%----------------------------------------------
\section{The Hessian $\Gamma^{(2)}_k$ for higher-derivative gravity}
\label{App:A}
%----------------------------------------------
\noindent
In four dimensions, the derivative expansion
 of $\Gamma_k^{\rm grav}[g]$ up to fourth order can be organized into 
 the five interaction monomials,
\be
\begin{array}{lll}
I_0 = \int \rmd^4x \sqrt{g} \, , \qquad & I_1 = \int \rmd^4x \sqrt{g} R \, , \\[1.1ex]
I_2 = \int \rmd^4x \sqrt{g} R^2 \, , \qquad & I_3 = \int \rmd^4x \sqrt{g} R_{\mu\nu}R^{\mu\nu} \, , \qquad & I_4 = \int d^4x \sqrt{g} E
\end{array}
\ee
where $E = R^2 - 4 R_{\mu \nu} R^{\mu \nu} + R_{\mu \nu \rho \sigma} \, R^{\mu \nu \rho \sigma}$ is the integrand 
of the Euler topological invariant, $\int d^4x \sqrt{g} E = 32 \pi^2 \chi$. Constructing the argument
of the traces entering into the FRGE requires the second variation of these invariants. In this context, we first note
that $I_4$ is a topological quantity, so that its variation with respect to the metric vanishes.
To obtain the hessians of the other invariants, 
we split $g_{\mu\nu} = \gb_{\mu\nu} + h_{\mu\nu}$, where $\gb_{\mu\nu}$ denotes a fixed background metric 
and $h_{\mu\nu}$ is an arbitrary fluctuation. The general expressions for these variations, valid for
an arbitrary background $\gb_{\mu\nu}$, can be found in \cite{Barth:1983hb} 
(see also \cite{Julve:1978xn,Avramidi:1985ki,Christensen:1978sc,deBerredoPeixoto:2003pj}). 
For our purposes, however, it suffices to consider these variations on backgrounds
 $\gb_{\mu\nu} = \gb^{\mathcal{E}}_{\mu\nu}$, where the index $\mathcal{E}$ indicates
that the background metric is a generic Einstein metric. These are metrics satisfying $\Rb_{\m\n}=\frac{\Rb}{d}\gb_{\m\n}$ (but not necessarily $\bar C_{\m\n\r\s}= 0$) and, using the contracted Bianchi identity, this condition also implies that $\bar D^\lambda \Rb_{\lambda\sigma\m\n} = 0$. For these spaces, the Hessians of $I_n$ then simplify considerably. At the two-derivative level, we obtain
\be
\begin{split}
\delta^2 I_0 = & \, \int_\mathcal{E} d^4x \sqrt{\gb} \left[ \tfrac{1}{4} h^2 - \tfrac{1}{2} h_{\mu\nu} h^{\mu\nu} \right] \, , \\
\delta^2 I_1 = & \, \int_\mathcal{E} d^4x \sqrt{\gb} \Big[
  \tfrac{1}{2} h^{\alpha \beta} \left[ \Db^2 - \tfrac{1}{2} \Rb \right] h_{\alpha \beta} + \Rb_{\alpha \mu \beta \nu} h^{\alpha \beta} h^{\mu\nu} \\
& \qquad \qquad \quad - \tfrac{1}{2} h \Db^2 h + 
h (\Db^\alpha  \Db^\beta h_{\alpha \beta}) + (\Db^\mu h_{\mu\alpha})(\Db_\nu h^{\nu\alpha})
 \Big] \, , 
\end{split}
\ee
while the variations of the four-derivative terms yield
\be\label{d2R2}
\begin{split}
\delta^2 I_2 = \int_\mathcal{E} d^4x \sqrt{\gb}   \Big\{&\,
2 h \left[ \Db^4 - \tfrac{1}{16} \Rb^2 \right] h
+ \Rb h^{\alpha \beta} \Db^2  h_{\alpha \beta}
+2 \Rb \, \Rb_{\alpha \mu \beta \nu } h^{\alpha \beta} h^{\mu \nu} 
+2 (\Db_\alpha \Db_\beta h^{\alpha \beta})^2 \\
& \quad 
+ (\Db_\alpha \Db_\beta h^{\alpha \beta}) \left[ -4 \Db^2 + \Rb \right]h  
+ 2 \Rb (\Db_\alpha h^{\alpha \beta})(\Db^\mu h_{\mu \beta})
\Big\} \, ,
\end{split}
\ee
and
\be\label{d2Ricci}
\begin{split}
\delta^2 I_3 = \int_\mathcal{E} d^4x & \, \sqrt{\gb} \Big\{ 
\tfrac{1}{2} h^{\alpha \beta} \left[ \Db^4 + \tfrac{1}{2} \Rb \Db^2  \right] h_{\alpha \beta}
+ \tfrac{1}{2} h \left[ \Db^4 - \tfrac{1}{4} \Rb \Db^2  - \tfrac{1}{8} \Rb^2 \right] h \\
& \,
- (\Db_\alpha \Db_\beta h^{\alpha \beta}) \left[ \Db^2 -  \tfrac{\Rb}{2} \right] h
+ (\Db_\alpha h^{\alpha \beta}) \left[ \Db^2 + \tfrac{3\Rb}{4} \right] (\Db^\mu h_{\mu \beta}) 
 \\
& \, + (\Db_\alpha \Db_\beta h^{\alpha \beta})^2 - 2 h_{\alpha \beta} \Rb^{\alpha \mu \nu \beta} \left[\Db^2 + \tfrac{1}{4} \Rb \right] h_{\mu \nu}
+ 2 h_{\alpha \beta} \Rb^{\alpha \lambda \beta \sigma } \Rb_{\lambda \mu  \sigma \nu} h^{\mu \nu}
\Big\} \, ,
\end{split}
\ee
respectively. Here, the bar denotes that the corresponding quantity is constructed from the background metric and $h = \gb^{\mu\nu} h_{\mu\nu}$.

A remarkable feature of these variations is that they can naturally be written in terms of second order minimal operators of
Lichnerowicz form \eqref{def:LL}. In particular, the four-derivative operators appearing in \eqref{d2Ricci} and \eqref{d2R2} factorize into squares of these (modified) Laplacians. Performing the TT-decomposition \eqref{TT-met} for the metric, a brief computation establishes
\be\label{varEH}
\begin{split}
\d^2 I_0 = & \int_\mathcal{E} d^4x \sqrt{\gb} \left\{ \tfrac{1}{8} h^2 - \tfrac{1}{2} h^{{\rm T} \mu\nu} h^{\rm T}_{\mu\nu} - \xi^\mu \Delta_{1L} \xi_\mu - \tfrac{1}{8} \sigma \left[ 3 \Delta_{0L} - \Rb \right] \Delta_{0L} \sigma \right\} \\
\d^2 I_1
= & \int_\mathcal{E} d^4x \sqrt{\gb} \Big\{ \, \tfrac{3}{16} h \D_{0L} h  - \tfrac{1}{2} h^{\rm T\,\m\n} \Big[  \D_{2L} + \tfrac{1}{2} \Rb \Big] h^{\rm T}_{\m\n} - \tfrac{1}{2} \Rb \, \xi^\n \D_{1L} \, \xi_\n \\
 &\, \qquad \qquad \quad 
 + \tfrac{1}{8} h \big[ 3\D_{0L}-\Rb \big]\D_{0L} \s + \tfrac{1}{16} \s \big[ \D_{0L}-\Rb \big] \big[3\D_{0L}-\Rb \big] \D_{0L}\s  \Big\} \, .
\end{split}
\ee
For the four-derivative terms, an analogous computation shows
\be\label{varR2}
\begin{split}
\delta^2 I_2  = \, \int_\mathcal{E} d^4x \sqrt{\gb}   \Big\{& \, \tfrac{3}{8}\,h\left[3\Delta_{0L}-\Rb\right]\Delta_{0L}\, h -\Rb\,h^{\alpha \beta{\rm T}}\,\Delta_{2L} \,h_{\alpha \beta}^{\rm T} \\&\quad\,
+ \tfrac{3}{8}\sigma \,\left[3\Delta_{0L}-\Rb\right]\Delta_{0L}^3 \,\sigma  
 + \tfrac{3}{4}h\left[3\Delta_{0L}-\Rb\right]\Delta_{0L}^2\,\sigma \Big\}\,,
\end{split}
\ee
and
\be\label{varRmn2}
\begin{split}
\delta^2 I_3 = \int_\mathcal{E}  d^4x  \sqrt{\gb} \Big\{& \,
\tfrac{1}{2} \, h^{{\rm T}\alpha \beta} \left[ \Delta_{2L} - \tfrac{1}{2} \Rb  \right] \Delta_{2L} \, h_{\alpha \beta}^{\rm T} 
+ \tfrac{1}{8} h \left[ 3 \Delta_{0L} - \Rb  \right] \Delta_{0L} h \\
& \, 
+ \tfrac{1}{8} \sigma \left[ 3 \Delta_{0L} - \Rb \right] \Delta_{0L}^3 \sigma 
+ \tfrac{1}{4} h \left[ 3 \Delta_{0L} - \Rb \right] \Delta_{0L}^2 \sigma 
\Big\} \, .
\end{split}
\ee
As a welcome side-effect, we also observe that the introduction of the Lichnerowicz-Laplacians
diagonalizes the transverse-traceless sector of the fluctuations. With respect to ``off-diagonal''
terms in the metric sector of $\Gamma^{(2)}_k[\gb]$, the $R^2 + C^2$-truncation has thus the same 
level of complexity as previous computations which included (polynomials of) the Ricci-scalar only 
and referred to a maximally symmetric background.  

In order to complete the construction of the operator traces, we now turn to the gauge-fixing
and ghost terms originating from \eqref{S:gf2}. For the higher-derivative action of Section \ref{sec:HD},
we thereby work with $\alpha = 0, \rho =0$. In this case, the TT-decomposition of $S^{\rm gf}$
yields
\be\label{Sgf}
S^{\rm gf} = - \frac{\beta}{2} \int d^4x \sqrt{\gb}
\left\{
\xi_\mu \big[ (\Delta_{1L} + \tfrac{\Rb}{4} ) \Delta_{1L}^2 \big] \xi^\mu
+ \sigma \big[ (\tfrac{3}{4} \Delta_{0L} - \tfrac{\Rb}{4} )^2 (\Delta_{0L} - \tfrac{\Rb}{4})\Delta_{0L}
\big] \sigma
\right\}
\ee
The ghost sector now contains, in addition to the usual (complex) $\Cb,C$-ghost fields, a third ghost \cite{Barth:1983hb} due to the two-derivative contribution
$(\det{\beta D^2})^{1/2}$. Introducing the complex-valued Grassmann fields $\bar{B}_\mu$, $B^\mu$ and the real field $b_\mu$ for the latter term, and TT-decomposing the ghost sector of the resulting action then leads to 
\be
\begin{split}
S_{\rm C-ghost}^{\rm quad} = & \, - \int d^4x \sqrt{\gb} \Big\{
\Cb^{\rm T}_\mu  \Delta_{1L} C^{{\rm T} \mu}
+ \tfrac{1}{2} \bar \eta \big[ 3 \Delta_{0L} - \Rb  \big] \Delta_{0L} \eta 
\Big\} \, , \\
S_{\rm B-ghost}^{\rm quad} = & \, -  \int d^4x \sqrt{\gb} \Big\{
\bar{B}_\mu^{\rm T} \big[ \Delta_{1L} + \tfrac{\Rb}{4} \big] B^{{\rm T}\mu}
+ \bar{B} \, \big[ \Delta_{0L} - \tfrac{\Rb}{4} \big] \Delta_{0L} \, B \\
& \qquad\qquad\, + \tfrac{1}{2} b_\mu^{\rm T} \big[ \Delta_{1L} + \tfrac{\Rb}{4} \big] b^{{\rm T}\mu}
+ \tfrac{1}{2} b \, \big[ \Delta_{0L} - \tfrac{\Rb}{4} \big] \Delta_{0L} \, b
\Big\} \, .
\end{split}
\ee
Note that, in the literature on higher-derivative gravity, the contribution of the $\bar{B},B$-ghost field is usually absorbed into the usual $\Cb,C$-ghost, hence the need of only a third (real) ghost. We prefer here to introduce a fourth ghost to clearly separate the higher-derivative contribution from the usual second order term. The two choices are of course equivalent. In the following, we impose a ``mode by mode'' cancellation between the gauge-degrees of freedom in the metric and the ghost sector \cite{Codello:2007bd}, which results in a precise cancellation of all the ``unphysical mode contributions'' to \eqref{Tr:ghost}. 

Finally, there are additional contributions to the flow equation arising from the Jacobi-determinants introduced via the TT-decomposition,
\be
\begin{split}
J_{\rm grav} = \Big( \det{}_{\rm (1T,0)}^\prime\left[ M^{(\mu, \nu)} \right] \Big)^{1/2} \,,\;
J_{\rm C-gh} =  J_{\rm B-gh} =\left( {\det}^\prime[\Delta_{0L}]\right)^{-1}\,,\;
J_{\rm b-gh} =  \left( {\det}^\prime[\Delta_{0L}]\right)^{1/2}\,.
\end{split}
\ee
Here, the primes indicate that the unphysical modes are left out from the determinants. Furthermore, $M^{(\mu, \nu)}$ is a $(d+1) \times (d+1)$-matrix differential operator whose first $d$ columns act on the transverse spin one fields $\xi_\mu$ and whose last column acts on the spin zero fields $\sigma$ and which reads
\be
M^{(\mu, \nu)} = \left[ 
\begin{array}{cc}
 2 \, g^{\mu \nu} \Delta_{1L} & -  \frac{R}{2} D^\mu \\
\frac{R}{2}D^\nu & \frac{3}{4} \Delta_{0L}^2 - \frac{R}{4}\Delta_{0L} \\
\end{array}
\right] \, .
\ee
In order to account for these contributions, we follow earlier works
\cite{Codello:2007bd,Machado:2007ea,Codello:2008vh} and introduce appropriate auxiliary fields
 so as to exponentiate these determinants via the Faddeev-Popov trick.
 The resulting ``auxiliary action'' then becomes
\be\label{Saux}
\begin{split}
S^{\rm aux} = \int \, d^4x \, \sqrt{g} \, \Big\{ & 
[\zeta^{{\rm T}}_\mu, \omega] \big[ M^{(\mu ,\nu)} \big]^\prime  [\zeta^{{\rm T}}_\nu, \omega]^{\rm T} +
 [\cb^{{\rm T} }_\mu, \bar{c}] \big[ M^{(\mu ,\nu)} \big]^\prime  [c^{\rm T}_\nu, c]^{\rm T} \\ &
+ \bar{s} \, \Delta_{0L}^\prime \, s  + \bar{t} \, \Delta_{0L}^\prime \,t
\,+\bar{\chi}\,\Delta_{0L}^\prime\,\chi +\tfrac{1}{2} \phi\,\Delta_{0L}^\prime\,\phi \Big\} \,  .
\end{split}
\ee
Here the gravitational sector contains the transverse ghost $\cb^{\rm T}_\mu, c^{{\rm T}\mu}$, a ``longitudinal'' Grassmann scalar $\cb, c$, a transverse vector $\zeta^{\rm T}_\mu$ and a real scalar $\omega$, while the ghost determinants are captured by the contribution of the complex scalar fields $s$, $\bar{s}$, $t$, $\bar{t}$, the complex Grassmann fields $\bar{\chi}$, $\chi$, and the real scalar field $\phi$.

We now have all the ingredients for constructing all the Hessians $\Gamma^{(2)}_k$ required in the r.h.s.\ of the flow equation. These are collected
in Table \ref{T.1}. 
\begin{table}[t!]
\begin{center}
\begin{tabular}{l|l}
Index & Hessian $\Gamma^{(2)}_k$  \\ \hline \hline
 $h^{\rm T} h^{\rm T}$ 
& $ 2 u_3 \D_{2L}^2 -  [2 u_2 R - \tfrac{1}{3} u_3 R + u_1] \D_{2L} - \tfrac{1}{2} u_1 R -  u_0$
\\[1.1ex]
$\xi_\m \xi^\m$
& $-\b[\D_{1L}+\tfrac{R}{4}]\D_{1L}^2-[u_1R+2u_0]\D_{1L}$
\\[1.1ex]
$h h$
& $ \tfrac{1}{8} \big[ 18 u_2 \D_{0L}^2 + 3 (u_1 - 2 u_2 R) \D_{0L} + 2 u_0 \big] $
 \\[1.1ex]
$\sigma \sigma$ 
& $ -\tfrac{\b}{16}\big[3\D_{0L}-R\big]^2\big[\D_{0L}-\tfrac{R}{4}\big]\D_{0L} + \tfrac{1}{8} \big[ 6 u_2 \D_{0L}^2 + u_1 \D_{0L} -u_1 R - 2 u_0 \big] \big[3\D_{0L}-R\big] \D_{0L}  $
 \\[1.1ex]
$h \sigma$ 
& $ \tfrac{1}{8} \big[u_1 + 6 \, u_2 \, \D_{0L}\big] \big[3 \D_{0L} - R\big] \D_{0L}$ \\[1.1ex] \hline
$\Cb^{\rm T}_\m C^{\rm T\m}$
& $ \b \D_{1L}$
 \\[1.1ex] 
$\eb \eta$
& $ \tfrac{\b}{8} [3 \D_{0L} - R] \D_{0L} $
\\[1.1ex]
$\bar B^T_\m B^{T\m}$
& $  -\b [ \D_{1L}+\tfrac{R}{4} ] $
\\[1.1ex]
$\bar B B$
& $ -\b [\D_{0L}-\tfrac{R}{4}] \D_{0L}$
\\[1.1ex]
$b^T_\m b^{T\m}$
& $ -\b  [\D_{1L}+\tfrac{R}{4}]$
\\[1.1ex]
$b b$
& $ -\b [\D_{0L}-\tfrac{R}{4}] \D_{0L}$
\\[1.1ex] \hline
$\zeta^T_\m \zeta^{T\m}$
& $ 4\D_{1L} $
 \\[1.1ex]
$\omega \omega$
& $ \tfrac{1}{2} [3\D_{0L}-R]\D_{0L} $
 \\[1.1ex]
$\cb^{\rm T}_\m c^{\rm T\m}$
& $2 \D_{1L}  $
\\[1.1ex]
$\cb c$
& $ \tfrac{1}{4} [3\D_{0L}-R]\D_{0L}$
\\[1.1ex]
$\bar{s} s$
& $ \D_{0L} $
 \\[1.1ex]
 $\phi \phi$
& $ \D_{0L} $
 \\[1.1ex] \hline \hline
\end{tabular}
\end{center}
\parbox[c]{\textwidth}{\caption{\label{T.1}{Matrix entries of the operator $\Gamma^{(2)}_k$ in the gravitational, ghost and auxiliary sector (separated by the horizontal lines), respectively. The elements are symmetric under the change of bosonic indices, while they acquire a minus sign when Grassmann-valued indices are swapped.}}}
\end{table}
The gravitational sector then arises from combining the contributions from the variations \eqref{varEH}, \eqref{varR2}, and \eqref{varRmn2} with the gauge-fixing action \eqref{Sgf},  reinstalling the corresponding coupling constants. These kernels contain all the information required for constructing the r.h.s. of the flow equation \eqref{Ftrace}. To cast the result into the form \eqref{Tr:grav} and \eqref{Tr:ghost} it is thereby useful to note that the $\s$-$h$-crossterm vanishes in the limit $\beta \rightarrow \infty$. Thus, the combined contribution from $\s$ and $h$ splits into the sum of the $hh$-trace \eqref{Tr:grav} and the contribution of the $\s\s$-part. The latter can be combined with the contribution of all the other scalar fields to give rise to the universal scalar trace $\cS_0$. Similarly, combining all contributions from transverse vectors leads to $\cS_{\rm 1T}$, which is also independent of the details of the gravitational action. Finally, the $h^{\rm T}h^{\rm T}$-sector produces the $\cS_{\rm 2T}$-trace. 

Lastly, we note that the derivation of the flow equation for the Einstein-Hilbert truncation proceeds in an entirely analogous manner. In this case, the gravitational sector arises from the contributions of \eqref{varEH} with the TT-decomposed gauge-fixing term \eqref{S:gf2} in the limit $\alpha\rightarrow\infty$, $\beta = 0$ and $\rho = 0$, with a similar vanishing of the $\s$-$h$-crossterm and decoupling of the $h$ and $\s$ traces. The ghost sector now contains only the C-ghosts, and the auxiliary sector consequently does not contain the $\phi$, $\chi$ and $t$ fields. Combining the $\s\s$ with all the other scalar field traces and the $\xi\xi$ with all the other transverse vector traces then results in the universal traces $\cS_0$ and $\cS_{\rm 1T}$, respectively, which are again given by \eqref{Tr:ghost}.  The remaining $\cS_{\rm 2T}$ and $\cS_{hh}$ traces can then be straightforwardly constructed from Table \ref{T.1} by setting the higher-derivative couplings to zero. This concludes our derivation of the gravitational sector of the flow equation \eqref{Ftrace}.

%----------------------------------------------
\section{Heat-kernel coefficients for Lichnerowicz Laplacians}
\label{App:B}
%----------------------------------------------
For evaluating the operator traces appearing in Section \ref{sec:HD}, we require
the heat-kernel expansion for the Lichnerowicz operators \eqref{def:LL}, evaluated at a
generic Einstein manifold, up to fourth order in the derivative expansion. In this appendix
we derive the corresponding coefficients starting from the early time heat-kernel expansion of a generic
two-derivative differential operator \cite{Gilkey:1995mj,Avramidi:2000bm} (see also \cite{Codello:2008vh} for a nice exposition
in the context of the FRG).
%----------------------------------------------
\subsection{Heat-kernel coefficients for unconstrained fields}
\label{App:B1}
%----------------------------------------------
In general, the early time heat-kernel expansion of a generic second
order differential operator $\Delta = -D^2 +\bQ$ takes the form
\be\label{Hexp}
\Tr\left[ \e^{\I t \Delta} \right] = \left( \frac{\I}{4 \pi t} \right)^{2} \int d^4x \sqrt{g}
\left\{ \tr \, a_0 - \I t \, \tr \, a_2 - t^2 \, \tr \, a_4 + \ldots \right\} \, ,
\ee
with the heat-kernel coefficients $a_{2k}$ given by \cite{Gilkey:1995mj}
\be
\begin{split}
a_0 = & \, \unit \, , \qquad 
a_2 = \, P \, , \\
a_4 = & \, \frac{1}{180} \left( R_{\m\n\alpha\beta} R^{\m\n\alpha\beta} - R_{\m\n}R^{\m\n} + D^2 R\right) \unit + \frac{1}{2} P^2 + \frac{1}{12} \cR_{\m\n} \cR^{\m\n} + \frac{1}{6} D^2 P \, .
\end{split}
\ee
Here, $D^2$ is the covariant Laplacian with respect to the (background) metric, $\bQ$ is a matrix-valued potential, $P = \tfrac{1}{6} R {\bf \unit} + \bQ$, $\cR_{\mu\nu} = 2 D_{[\mu} D_{\nu]}$ is the commutator of the covariant derivatives, and $\tr$ denotes a trace with respect to the spin-indices of the fields on which $\Delta$ acts. For the purpose of 
this paper, we have to evaluate $\tr_s \, a_{2k}$ for scalars ($s=0$), vectors ($s=1$), and symmetric tensors $(s=2)$. In the latter two cases, the $\tr_s$ are defined as 
\be
\tr_{1} a_{2k} = g^{\m\n} [a_{2k}]_{(\m\n)} \, , \qquad \tr_{2} a_{2k} = g^{\mu \rho} g^{\n \sigma} [a_{2k}]_{(\m\n)(\rho\sigma)} \, ,
\ee
respectively. The matrices $\cR_{\mu\nu} \cR^{\mu\nu}$ are trivial for the scalar case, whereas for vectors and tensors they respectively read
\be
\begin{split}
\left[ \cR_{\alpha\beta} \cR^{\alpha\beta} \right]_{\m\n} = & \, - R_{\alpha \beta\gamma \mu} R^{\alpha \beta\gamma}{}_{\nu} \, , \quad \\
\left[ \cR_{\alpha\beta} \cR^{\alpha\beta} \right]_{\m\n\rho\sigma} = & \,
- R_{\alpha \beta\gamma \mu} R^{\alpha \beta\gamma}{}_{\rho} \, g_{\nu \sigma}
- R_{\alpha \beta\gamma \nu} R^{\alpha \beta\gamma}{}_{\sigma} \, g_{\mu \rho}
+ 2 R_{\alpha \beta\mu\rho} R^{\alpha \beta}{}_{\nu\sigma} \, .
\end{split}
\ee

The differential operators appearing in the traces \eqref{Tr:grav} and \eqref{Tr:ghost} are the Lichnerowicz operators
\eqref{def:LL}, i.e., second order differential operators with matrix-potentials
\be
\bQ_0 = 0 \, , \qquad 
[\bQ_1]_{\m\n} = \frac{1}{4} g_{\m\n} R \, , \qquad
[\bQ_2]_{\mu\nu\alpha\beta} = 2 R_{\mu\alpha\nu\beta} \,.
\ee
Their heat-kernel coefficients on a generic four-dimensional Einstein manifold without boundary 
can be obtained by substituting these potentials into the expressions for the generic heat-kernel expansion. Evaluating the spin-traces, we obtain
\be
\begin{array}{lll}
\tr_0 a_0 = 1 \, , \qquad &  
\tr_0 a_2 = \frac{1}{6} R \, , \qquad & 
\tr_0 a_4 = \frac{1}{180} R_{\m\n\alpha\beta} R^{\m\n\alpha\beta}  + \frac{1}{80}R^2 \, , 
\\[1.2ex]
\tr_1 a_0 = 4 \, , & 
\tr_1 a_2 = \frac{5}{3} R \, , & 
\tr_1 a_4 = -\frac{11}{180} R_{\m\n\alpha\beta} R^{\m\n\alpha\beta}  
+ \frac{41}{120}R^2 \, , 
\\[1.2ex]
\tr_{2} a_0 = 10 \, , &
\tr_{2} a_2 = \frac{2}{3} R \, , &
\tr_{2} a_4 = \frac{19}{18} R_{\m\n\alpha\beta} R^{\m\n\alpha\beta}  
- \frac{1}{24}R^2 \, . 
\end{array}
\ee
This result completes the heat-kernel expansion for unconstrained fields.
%  
%----------------------------------------------
\subsection{Heat-kernel coefficients for fields with differential constraints}
\label{App:B2}
%----------------------------------------------
In order to apply the early-time heat-kernel expansion to the operator traces \eqref{Tr:grav}, 
and \eqref{Tr:ghost} the heat-kernel coefficients for the unconstrained fields given in the last subsection 
must be converted into the expansion coefficients for the transverse vectors (1T) and transverse-traceless
symmetric matrices (2T) entering into the TT-decomposition. 

In the decomposition of a vector field into its transverse and longitudinal
parts,
\be\label{TT-vec}
A_\mu = A_\mu^{\rm T} + D_\mu \Phi \; , \qquad D^\mu A_\mu^{\rm T} = 0 \, ,
\ee
the spectra of $D_\mu \Phi$ and $\Phi$ are related by 
\be
\Delta_{1L} D_\mu \Phi = D_\mu (\Delta_{0L} - \tfrac{1}{2} R ) \Phi \, , 
\ee
and the constant mode in $\Phi$ does not contribute to $A_\mu$.   
Thus, the decomposition of the $s=1$-trace takes the form 
\be\label{h1T}
\Tr_{\rm 1T}\left[ \e^{\I t \D_{1L}} \right] 
= \Tr_{1}\left[ \e^{\I t \D_{1L}} \right]
- \Tr_{0}\left[ \e^{\I t (\D_{0L} - R/2)} \right]
+ \e^{-\I t R/2} \, .  
\ee
where the last term removes the constant $\Phi$-mode from the $s=0$-trace. A similar
argument applies to the TT-decomposition of the symmetric tensor,
\be \label{TT-met}
h_{\mu\nu} = h_{\mu\nu}^{\rm T} + 2 D_{(\mu} \xi_{\nu)} + D_\mu D_\nu \sigma +\frac{1}{4} g_{\mu\nu} \Delta_{0L} \sigma + \frac{1}{4} g_{\mu\nu} h\,,
\ee
where the components appearing on the RHS of this decomposition are subject to the constraints
\be
g^{\mu \nu} \, h_{\mu\nu}^{\rm T} = 0 \, , \quad D^\mu h_{\mu\nu}^{\rm T} = 0
\, , \quad D^\mu \xi_\mu = 0 \, , \quad h = g_{\mu \nu} h^{\mu \nu} \, .
\ee
In this case, one can use
\be\label{id1}
\begin{split}
& \,\Delta_{2L} D_\mu \xi_\nu  =  D_\mu \Delta_{1L} \xi_\nu\,, \\
& \,\Delta_{2L} \left[ D_\mu D_\nu +\tfrac{1}{4} g_{\mu\nu} \Delta_{0L} \right] \sigma  =  \left[ D_\mu D_\nu +\tfrac{1}{4} g_{\mu\nu} \Delta_{0L} \right] \left[ \Delta_{0L} - \tfrac{R}{2} \right] \sigma  \, , \\
& \, \Delta_{2L} g_{\mu\nu} h  =  g_{\mu\nu}\left[\Delta_{0L} - \tfrac{R}{2}\right] h \, ,
\end{split}
\ee
to relate the spectrum of $\Delta_{2L}$ to the ones of the vector and scalar fields. Furthermore,
\eqref{TT-met} indicates that  the constant mode in $\sigma$, 
scalars subject to $\left[D_\mu D_\nu  + \tfrac{1}{4} g_{\mu\nu} \Delta_{0L} \right] \sigma = 0$, 
and transverse vectors satisfying $D_{(\mu} \xi_{\nu)} = 0$ do not
contribute to $h_{\mu\nu}$, so that the corresponding modes have to be removed from 
the decomposed spectrum. By contracting the last two equations with
$D^\nu$, one can show that these are eigenmodes of  $\Delta_{0L}$ and $\Delta_{1L}$
with eigenvalues 
$\Lambda_{0L} = 0$, $\Lambda_{0L} = \frac{R}{3}$, and $\Lambda_{1L} = 0$,
respectively.\footnote{For a spherical background, these coincide with the two lowest eigenmodes of $-D^2$ acting on scalars and the lowest eigenmode of
$-D^2$ acting on vector fields \cite{Lauscher:2002mb,Codello:2008vh}.}
The multiplicity of the latter two is given by
the number of  Killing vectors $n_{\rm KV}$ and conformal Killing vectors $n_{\rm CKV}$
of the background. Taking into account \eqref{h1T}, the operator trace
for transverse-traceless tensors field can then be expressed in terms of traces over unconstrained fields
\be\label{h2T}
\begin{split}
\Tr_{\rm 2T}\left[ \e^{\I t \D_{2L}} \right] = & \, 
\Tr_{2}\left[ \e^{\I t \D_{2L}} \right] 
- \Tr_{1}\left[ \e^{\I t \D_{1L}} \right]
- \Tr_{0}\left[ \e^{\I t (\D_{0L} -R/2)} \right] 
 +  n_{\rm KV}+ n_{\rm CKV} \e^{-\I t R/6 }  \, .
\end{split}
\ee
In the following, we will assume that our background is generic, in the sense that its metric does not admit
Killing or conformal Killing vectors. 

From eqs.\ \eqref{h1T} and \eqref{h2T} it is then straightforward to compute the heat-kernel coefficients for Lichnerowicz
Laplacians acting on transverse vectors and transverse traceless symmetric matrices.  
For a generic Einstein background, these read
\be\label{d4heat}
\begin{array}{lll}
\tr_0 a_0 = 1 \, , \quad & 
\tr_0 a_2 = \frac{1}{6} R \, , \quad &
\tr_0 a_4 = \frac{1}{180} \, R_{\m\n\alpha\beta} R^{\m\n\alpha\beta} + \frac{1}{80} \, R^2 \, , \\[1.1ex]
\tr_{\rm 1T} a_0 = 3 \, , \quad & 
\tr_{\rm 1T} a_2 = R \, , \quad &
\tr_{\rm 1T} a_4 = - \frac{1}{15} \, R_{\m\n\alpha\beta} R^{\m\n\alpha\beta} + \frac{29}{240} \, R^2 \, , \\[1.1ex]
\tr_{\rm 2T} a_0 = 5 \, , \quad & 
\tr_{\rm 2T} a_2 = - \frac{5}{3} R \, , \quad &
\tr_{\rm 2T} a_4 = \frac{10}{9} \, R_{\m\n\alpha\beta} R^{\m\n\alpha\beta} - \frac{29}{48} \, R^2 \, .\\
\end{array}
\ee
These coefficients are the key ingredient for evaluating the operator traces 
 \eqref{Tr:grav} and \eqref{Tr:ghost}
and constitute the main result of this appendix.

%----------------------------------------------------------------------------
\section{Operator traces and $\beta$-functions} 
\label{App:C}
%----------------------------------------------------------------------------
In this appendix, we evaluate the operator traces appearing
on the r.h.s.\ of eq.\ \eqref{Ftrace}. We start with reviewing
some general properties and definitions before computing 
the traces entering into our truncations explicitly.
%----------------------------------------------------------------------------
\subsection{General trace technology}
\label{App:C1}
%----------------------------------------------------------------------------
The key observation for evaluating the traces entering
\eqref{Ftrace} is that they contain only minimal second order
differential operators, which commute with all other elements 
(like the curvature scalars) inside the trace. Their projection
onto the truncation subspace can then be found using the
heat-kernel coefficients for constrained fields given in \eqref{d4heat}. 
Here, the key formula is
\be\label{heat:master}
{\rm Tr}[W(\Delta_{iL})] = \frac{1}{(4\pi)^2} \int \rmd^4x \sqrt{\gb} \left\{
Q_2[W] {\rm tr}_i \, a_0 + Q_1[W] {\rm tr}_i \, a_2 + Q_0[W] {\rm tr}_i \, a_4 
+ \ldots \right\} \, ,  
\ee
where $W(z)$ is a smooth function whose argument has been replaced by the Lichnerowicz operators and where the dots indicate higher-derivative terms at order six and higher,
which are outside our truncation subspace. The functionals $Q_n[W], n \ge 0$ 
are defined as 
\be
\begin{split}
Q_n[W] = & \frac{1}{\Gamma(n)} \int_0^\infty dz \, z^{n-1} W(z) \; , \; n > 0 \, , \qquad
Q_0[W] =  W(0) \, .
\end{split}
\ee
In order to construct the $\beta$-functions for the dimensionless couplings, it is useful
to convert the $Q_n[W]$ into standardized dimensionless threshold functions,
\be\label{defPhi}
\begin{split}
\Phi^p_n(w) := & \, \frac{1}{\Gamma(n)} \int^\infty_0 \rmd z \, z^{n-1} \, \frac{R^{(0)}(z) - z R^{(0)\prime}(z)}{(z + R^{(0)}(z)+w)^p} \, , \\
\Pt^p_n(w) := & \, \frac{1}{\Gamma(n)} \int^\infty_0 \rmd z \, z^{n-1} \, \frac{R^{(0)}(z)}{(z + R^{(0)}(z)+w)^p} \, , 
\end{split}
\ee
and their generalizations for higher-derivative theories,
\be\label{defUps}
\begin{split}
\Upsilon^p_{n,m}(u,v,w) := & \, \frac{1}{\Gamma(n)} \int_0^\infty \! \rmd z \, z^{n-1} \,
\frac{\left(z + R^{(0)}(z) \right)^m \, \left( R^{(0)}(z) - z R^{(0) \prime}(z) \right) }{\left( u \left(z + R^{(0)}(z) \right)^2 + v \left(z + R^{(0)}(z) \right) + w \right)^p}
\, , \\
\tilde{\Upsilon}^p_{n,m,l}(u,v,w) := & \, \frac{1}{\Gamma(n)} \int_0^\infty \! \rmd z \, z^{n-1} \,
\frac{\left(z + R^{(0)}(z) \right)^m \, \left(2z + R^{(0)}(z) \right)^l \, R^{(0)}(z) }{\left( u \left(z + R^{(0)}(z) \right)^2 + v \left(z + R^{(0)}(z) \right) + w \right)^p} \, , 
\end{split}
\ee
defined for $n>0$. For the particular functions $W$ occurring in this paper, these relations 
are 
\be\label{Phigen}
Q_n\left[\tfrac{\p_t (g_k R_k)} {(2 g_k)(P_k + c_k)^{p}} \, \right] = k^{2(n-p+1)} \left( \Phi^p_n(c_k/k^2) + \tfrac{1}{2} \p_t \ln(g_k) \, \tilde{\Phi}^p_n(c_k/k^2)  \right) \, ,  \quad n > 0 , 
\ee
in the case of the Einstein-Hilbert truncation, while for the $R^2 + C^2$-truncation
we additionally have
\be\label{Qgen}
\begin{split}
Q_n & \left[ \frac{\p_t \left(g_k (P_k^2 - \Delta^2) + \gt_k R_k \right)}{(u_k P_k^2 + v_k P_k + w_k)^p}\right] \\ 
& \, \quad = k^{2(n-2p+2)} \left\{ \p_t g_k \, \tilde{\Upsilon}^p_{n,0,1}(u_k,v_k/k^2,w_k/k^4) + 4 g_k \Upsilon^p_{n,1}(u_k,v_k/k^2,w_k/k^4) \right\} 
\\ & \, \qquad
+ k^{2(n-2p+1)} \left\{ \p_t \gt_k \, \tilde{\Upsilon}^p_{n,0,0}(u_k,v_k/k^2,w_k/k^4) + 2 \gt_k \Upsilon^p_{n,0}(u_k,v_k/k^2,w_k/k^4) \right\} \, .
\end{split}
\ee

All numerical evaluations require a particular choice of the cutoff-function $R^{(0)}(z)$. For the purpose of this paper, we will restrict ourselves to the use of the optimized cutoff \cite{Litim:2001up} where $R^{(0)}_{\rm opt} = (1-z) \theta(1-z)$. The main virtue of this choice of cutoff is, that all the integrals appearing in the (generalized) threshold functions can be carried out analytically. In particular,
\be\label{defPhiOpt}
\Phi^p_n(w) = \frac{1}{\Gamma(n+1)} \, \frac{1}{(1+w)^p} \, , \quad \Pt^p_n(w) = \frac{1}{\Gamma(n+2)} \, \frac{1}{(1+w)^p} \, , \quad \varphi \equiv \p_t \ln(R_k) = 2.
\ee
Similarly, the generalized threshold functions \eqref{defUps} become
\be\label{upsopt}
\begin{split}
\Upsilon^p_{n,m}(u,v,w) =  & \, \frac{1}{\Gamma(n+1)} \, \frac{1}{\left( u + v + w \right)^{p}} \, , \\
\tilde{\Upsilon}^p_{n,m;l}(u,v,w) = & \, \frac{(-1)^n}{\Gamma(n)} \, \frac{ \beta(-1,n,l+1) + \beta(-1,n+1,l+1) }{ (u + v+w)^{p}} \, .
\end{split}
\ee
Here, $\beta(-1,n,l)$ denotes the incomplete beta function. For fixed values $n,l$, these become constants. 
This property leads to considerable simplifications in the analysis of the corresponding $\beta$-functions. 
%
%----------------------------------------------------------------------------
\subsection{Evaluation of the traces}
\label{App:C2}
%----------------------------------------------------------------------------
%
The evaluation of the operator traces proceeds by expanding the arguments in a Taylor series in $R$ around $R=0$, keeping terms up to $R^2$ only. The operator
traces appearing as ``expansion coefficients'' can then be evaluated with the heat-kernel techniques introduced in the last subsection, c.f. eq.\ \eqref{heat:master}.
In particular, the functionals $Q_n[W]$ arising in these cases are of the form \eqref{Phigen} or \eqref{Qgen}, so that expressing them in terms of the generalized threshold functions is rather straightforward. We will now give the results for this evaluation for the various traces appearing in the main part of the paper, projecting the resulting RG flow onto the subspaces spanned by our truncations.

The functions $Q_n[W]$ featuring in the universal traces $\cS_0$ and $\cS_{\rm 1T}$ and the matter trace \eqref{mattr} are a special case of \eqref{Phigen} with $c_k = 0$ and $g_k$ a $k$-independent constant, which we can set to one. Following the strategy outlined above, the evaluation of the matter trace \eqref{mattr} results in
\be\label{mattercont}
\cS_{\rm matter} = \frac{1}{(4\pi)^2} \int d^4x \sqrt{g} \left[ k^4 \Phi^1_2 + \frac{1}{6} k^2 \Phi^1_1 R + \frac{\varphi}{2} (\frac{1}{180} R_{\mu\nu\rho\sigma} R^{\mu\nu\rho \sigma} + \frac{1}{80} R^2 ) \right] \, ,
\ee
while the  expansion of the universal traces \eqref{Tr:ghost} results in  
\be\label{eq:TRuni}
\begin{split}
\cS_0 = & - \tfrac{1}{(4\pi)^2} \int \rmd^4x\sqrt{g}
\left[ k^4 \Phi^1_2 + \tfrac{1}{6} (\Phi^1_1 + 2 \Phi^2_2 ) k^2 R  
+ (\tfrac{\varphi}{160} + \tfrac{1}{18} \Phi^2_1 + \tfrac{1}{9} \Phi^3_2 ) R^2 + 
\tfrac{\varphi}{360} R_{\mu\nu\rho\sigma} R^{\mu\nu\rho\sigma}
\right], \\
\cS_{\rm 1T} = & - \tfrac{1}{(4\pi)^2} \int \rmd^4x\sqrt{g}
\left[ 3 k^4 \Phi^1_2 + \Phi^1_1 k^2 R + \tfrac{29}{480} \varphi R^2 - \tfrac{\varphi}{30} R_{\mu\nu\rho\sigma} R^{\mu\nu\rho\sigma} \right] \, .
\end{split}
\ee
Here, all the $\Phi^p_n$ are evaluated at zero argument, and for the optimized cutoff they are trivially obtained from \eqref{defPhiOpt}.

The non-universal traces entering into the computation of the $\beta$-functions in the Einstein-Hilbert truncation \eqref{NU-EH}
can be evaluated along the same lines, utilizing the general relation \eqref{Phigen}. The results then read
\be
\begin{split}
\cS_{\rm hh} =  \frac{1}{(4\pi)^2} \int \rmd^4x\sqrt{g} &
 \Big[ 
k^4 \left( \Phi^1_2(-4\lambda/3) + \tfrac{1}{2} \p_t \ln(u_1) \tilde{\Phi}^1_2(-4\lambda/3) \right) \\ & \qquad
 + \tfrac{1}{6} k^2 R \left( \Phi^1_1(-4\lambda/3) + \tfrac{1}{2} \p_t \ln(u_1) \tilde{\Phi}^1_1(-4\lambda/3) \right)
 \Big] \, ,
\end{split}
\ee
and
\be
\begin{split}
\cS_{\rm 2T} = & \frac{1}{(4\pi)^2} \int \rmd^4x\sqrt{g} 
\Big[ 
5 k^4 \left( \Phi^1_2(-2\lambda) + \tfrac{1}{2} \p_t \ln(u_1) \tilde{\Phi}^1_2(-2\lambda) \right) \\ & \quad
- 5 k^2 R \left( \tfrac{1}{3} \Phi^1_1(-2\lambda) + \tfrac{1}{2} \Phi^2_2(-2\lambda) 
+\tfrac{1}{2} \p_t \ln(u_1) \left( \tfrac{1}{3} \tilde{\Phi}^1_1(-2\lambda) + \tfrac{1}{2} \tilde{\Phi}^2_2(-2\lambda) \right)
\right)
\Big] \, , 
\end{split}
\ee
respectively.

The evaluation of the non-universal traces \eqref{Tr:grav} appearing in the $R^2+C^2$ truncation,  on the other hand, is slightly more involved.
Applying \eqref{Qgen}, 
the expansion of $\cS_{hh}$ on the truncation subspace takes the form
\be\label{eq:TRh}
\cS_{hh} = \frac{1}{2(4\pi)^2} \int d^4x \sqrt{g}
\left\{ k^4 \, C_1 + C_2 k^2 R + \frac{1}{180} C_3 R_{\alpha \beta\mu\nu} \, R^{\alpha \beta\mu\nu} + C_4 R^2 \right\} \, .
\ee
The dimensionless coefficients $C_i$ can be readily expressed in terms of the generalized threshold functions \eqref{defUps} with arguments $\U^p_{n,m} = \U^p_{n,m}(6g_2,g_1, 2/3 g_0)$, $\tU^p_{n,m,l} = \tU^p_{n,m,l}(6g_2,g_1, 2/3 g_0)$ and read
\be\label{Ccoeff}
\begin{split}
C_1 = & \,  24 \, g_2 \Upsilon^1_{2,1} + 2 g_1 \U^1_{2,0}
+ 6 \dg_2 \tU^1_{2,0,1} + (2 g_1 + \dg_1) \tU^1_{2,0,0} \, , \\
C_2 = & \, 4 g_2 \left( 12 g_2 \U^2_{2,2} - \U^1_{2,0} + \U^1_{1,1} + g_1 \U^2_{2,1} \right)  + \tfrac{1}{3} g_1 \U^1_{1,0} \\ &
- \dg_2 \big( 2 \tU^1_{2,0,0} - 12 g_2 \tU^2_{2,1,1} - \tU^1_{1,0,1} \big) 
+ (2 g_1 + \dg_1) (2 g_2 \tU^2_{2,1,0} + \tfrac{1}{6} \tU^1_{1,0,0}) \, , \\
C_3 = & \, \frac{(12 g_2 + g_1) \varphi + 6 \dg_2 + 2 g_1 + \dg_1}{6 g_2 + g_1 + \tfrac{2}{3} g_0} \, , \\
C_4 = & \,   g_2 \big\{ 96 g_2^2 \U^3_{2,3} - 8 g_2 ( \U^2_{2,1} - \U^2_{1,2} - g_1 \U^3_{2,2} ) + \tfrac{2}{3} (g_1 \U^2_{1,1} - \U^1_{1,0} ) \big\} \\
& \, + \dg_2 \big\{ 24 g_2^2 \tU^3_{2,2,1} - 2 g_2 ( 2 \tU^2_{2,1,0} - \tU^2_{1,1,1}) - \tfrac{1}{3} \tU^1_{1,0,0} \big\} \\& \,
 + g_2 \, (2 g_1 + \dg_1) \, (4 g_2 \tU^3_{2,2,0} + \tfrac{1}{3} \tU^2_{1,1,0} ) + \tfrac{1}{80} C_3 .
\end{split}
\ee
Here, we applied $R^{(0)}(0) = 1$ to simplify $C_3$ and expressed the resulting coefficients
in terms of the dimensionless coupling constants \eqref{dimless}.

The projection of $\cS_{\rm 2T}$ proceeds in a similar fashion. In this case all threshold functions appear with arguments
$\U^p_{n,m} = \U^p_{n,m}(2g_3,-g_1, -g_0)$, $\tU^p_{n,m,l} = \tU^p_{n,m,l}(2g_3,-g_1, -g_0)$. Parameterizing
\be\label{eq:TR2T}
\cS_{\rm 2T} = \frac{1}{2(4\pi)^2} \int d^4x \sqrt{g}
\left\{  k^4 \tC_1 + \tC_2 k^2 R + \frac{10}{9} \tC_3 R_{\alpha \beta\mu\nu} \, R^{\alpha \beta\mu\nu} + \tC_4 R^2 \right\} \, ,
\ee
the coefficients $\tC_i$ appearing in the trace expansion are
\be\label{tCcoeff}
\begin{split}
\tC_1 = & \, 40 g_3 \U^1_{2,1} - 10 g_1 \U^1_{2,0} + 10 \dg_3 \tU^1_{2,0,1} - 5 (2 g_1 + \dg_1) \tU^1_{2,0,0} \, , \\
\tC_2 = & 10 g_\flat \big( 4 g_3 \U^2_{2,2} -  \U^1_{2,0} -  g_1 \U^2_{2,1} \big)
 + 5 g_1 \big( 4 g_3 \U^2_{2,1} + \tfrac{2}{3} \U^1_{1,0} - g_1 \U^2_{2,0} \big) - \tfrac{40}{3} g_3 \U^1_{1,1} \\
& + 5 \dg_3 \big( 2 g_\flat \tU^2_{2,1,1} + g_1 \tU^2_{2,0,1} - \tfrac{2}{3} \tU^1_{1,0,1} \big) 
- 5 (2 g_1 + \dg_1) (g_\flat \tU^2_{2,1,0} - \tfrac{1}{3} \tU^1_{1,0,0} + \tfrac{1}{2} g_1 \tU^2_{2,0,0} )\\ & 
- 5\dg_\flat \tU^1_{2,0,0} \, , \\
\tC_3 = & \, \frac{(4 g_3 - g_1) \varphi + 2 \dg_3 - (2 g_1 + \dg_1)}{2 g_3 - g_1 - g_0} \, , \\
\tC_4 = & 5 g_\flat g_1 \left\{ 8 g_3\U^3_{2,2} - \U^2_{2,0} - 2 g_1 \U^3_{2,1} + \tfrac{2}{3} \U^2_{1,1} \right\} + \frac{10}{3} g_\flat \left\{ \U^1_{1,0} - 4 g_3 \U^2_{1,2} \right\} - \tfrac{20}{3} g_1 g_3 \U^2_{1,1} \\
& + 5 g_\flat^2 \left\{ 8 g_3 \U^3_{2,3} -2 \U^2_{2,1} - 2 g_1 \U^3_{2,2} \right\} 
+ 5 g_1^2 \left\{2 g_3 \U^3_{2,1} +\tfrac{1}{3} \U^2_{1,0} - \half g_1 \U^3_{2,0}   \right\} 
\\ 
& + 5 \, (2 g_1 + \dg_1) \left\{ g_\flat \, ( \tfrac{1}{3} \tU^2_{1,1,0} - g_1 \tU^3_{2,1,0} - g_\flat \tU^3_{2,2,0})  +   \tfrac{1}{6} g_1 \tU^2_{1,0,0} - \tfrac{1}{4} g_1^2 \tU^3_{2,0,0} \right\} \\
& + 5 \dg_3 \left\{ g_\flat ( 2 g_\flat \tU^3_{2,2,1} + 2 g_1 \tU^3_{2,1,1} - \tfrac{2}{3} \tU^2_{1,1,1}) - \tfrac{1}{3} g_1 \tU^2_{1,0,1} + \tfrac{1}{2} g_1^2 \tU^3_{2,0,1} \right\} \\
& - 5 \dg_\flat \left\{ g_\flat \tU^2_{2,1,0} + \half g_1 \tU^2_{2,0,0} - \tfrac{1}{3} \tU^1_{1,0,0} \right\}
- \tfrac{29}{48} \tC_3 \, .  
\end{split}
\ee
Note that the generalized threshold functions entering into $C_i$ and $\tC_i$ depend on {\it different arguments}. 

Substituting these expressions into the generic form of the flow equation \eqref{Ftrace} and comparing the coefficients on
the left and the right-hand-side then gives rise to the $\beta$-functions \eqref{betaEH} 
for the Einstein-Hilbert case and \eqref{beta} for the $C^2 + R^2$-truncation, respectively.

\end{appendix}

\end{document}